\documentclass{article}
\usepackage[hyphens]{url}
\usepackage[utf8]{inputenc}
\usepackage{graphicx}
\usepackage{authblk}
\usepackage[english]{babel}
\usepackage{setspace}
\usepackage{fancyhdr}
\usepackage{array}
\usepackage[margin=1in]{geometry}
\usepackage{float}
\usepackage{amsmath}
\usepackage{amssymb}
\usepackage{bbm}
\usepackage{bm}
\usepackage{scalerel,stackengine}
\usepackage[final]{pdfpages}
\usepackage{longtable}
\usepackage{chngcntr}
\usepackage{placeins}
\usepackage{microtype}
\usepackage{tikz}
\usepackage{makecell}
\usepackage{hyperref}
\usepackage{subcaption}
\usepackage{rotating}
\usepackage{soul}
\usepackage{tabularx} 
\usepackage{pdfpages}
\usepackage[square,numbers]{natbib}
\hypersetup{
    colorlinks = true,
    citecolor = {blue},
    urlcolor = {blue},
    menucolor = {blue},
    linkcolor = {blue}
}

\makeatletter
\patchcmd{\@maketitle}{\LARGE \@title}{\fontsize{18}{20}\selectfont\textbf{\@title}}{}{}
\makeatother

\title{On model-based time trend adjustments in platform trials with non-concurrent controls}

\author[1]{Marta Bofill Roig}
\author[1]{Pavla Krotka}
\author[2]{Carl-Fredrik Burman}
\author[3,4]{Ekkehard Glimm}
\author[5,6,7]{Stefan M. Gold}
\author[8]{Katharina Hees}
\author[9,10]{Peter Jacko}
\author[1]{Franz Koenig}
\author[3]{Dominic Magirr}
\author[11]{Peter Mesenbrink}
\author[12]{Kert Viele}
\author[ ]{Martin Posch$^{1,}$\thanks{martin.posch@meduniwien.ac.at}, \ on behalf of EU-PEARL (EU Patient-cEntric clinicAl tRial pLatforms) Consortium}

\affil[1]{Center for Medical Statistics, Informatics and Intelligent Systems, Medical University of Vienna, Vienna, Austria}
\affil[2]{Statistical Innovation, Data Science \& Artiffcial Intelligence, AstraZeneca, Gothenburg, Sweden}
\affil[3]{Advanced Methodology and Data Science, Novartis Pharma AG, Basel, Switzerland}
\affil[4]{Institute of Biometry and Medical Informatics, University of Magdeburg, Magdeburg, Germany}
\affil[5]{Klinik für Psychiatrie und Psychotherapie, Campus Benjamin Franklin, Charité -- Universitätsmedizin Berlin, Berlin, Germany}
\affil[6]{Medizinische Klinik m.S. Psychosomatik, Campus Benjamin Franklin, Charité -- Universitätsmedizin Berlin, Berlin, Germany}
\affil[7]{Institut für Neuroimmunologie und Multiple Sklerose (INIMS), Zentrum für Molekulare Neurobiologie, Universitätsklinikum Hamburg Eppendorf, Hamburg, Germany}  
\affil[8]{Section of Biostatistics, Paul-Ehrlich-Institut, Langen, Germany}
\affil[9]{Berry Consultants, UK.}
\affil[10]{Lancaster University, Lancaster, UK}
\affil[11]{Analytics Global Drug Development, Novartis Pharmaceuticals Corporation, One Health Plaza, East Hanover, NJ, USA.}
\affil[12]{Berry Consultants, USA}

\date{}         
\begin{document}

\maketitle

\begin{abstract}
\textbf{Background} Platform trials can evaluate the efficacy of several experimental treatments compared to a control. The number of experimental treatments is not fixed, as arms may be added or removed as the trial progresses. Platform trials are more efficient than independent parallel group trials because of using shared control groups.  
However, for a treatment entering the trial at a later time point, the control group is  
divided into concurrent controls, consisting of patients randomised to control when that treatment arm is in the platform, and non-concurrent controls, patients randomised before. Using non-concurrent controls in addition to concurrent controls can improve the trial’s efficiency by increasing power and reducing the required sample size, but can introduce bias due to time trends. 

\textbf{Methods} 
We focus on a platform trial  with two treatment arms and a common control arm. Assuming that the second treatment arm is added at a later time, we assess  the robustness of recently proposed model-based approaches  to adjust for time trends when utilizing non-concurrent controls. In particular, we consider approaches where time trends are modeled either as linear in time or as a step function, with steps at time points where treatments enter or leave the platform trial.  For trials with continuous or binary outcomes, we investigate the type 1 error rate and power of testing the efficacy of the newly added arm, as well as the bias and root mean squared error of treatment effect estimates under a range of scenarios. In addition to scenarios where time trends are equal across arms, we investigate settings with different time trends or time trends that are not additive in the scale of the model. 

\textbf{Results} A step function model, fitted on data from all treatment arms, gives increased power while controlling the type 1 error, as long as the time trends are equal for the different arms and additive on the model scale. This holds even if the shape of the time trend deviates from a step function when patients are allocated to arms by block randomisation. However, if time trends differ between arms or are not additive to treatment effects in the scale of the model, the type 1 error  rate may be inflated.

\textbf{Conclusions}
The efficiency gained by using step function models to incorporate non-concurrent controls can outweigh   potential risks of biases, especially in settings with small sample sizes. Such biases may arise if the model assumptions of equality and additivity of time trends  are not satisfied. However, the specifics of the trial, scientific plausibility of different time trends, and robustness of results should be carefully considered. 
\end{abstract}



\section{Background} 

Platform trials are  multi-arm trials that allow new experimental treatment arms to enter and leave the trial over time \cite{Woodcock2017,Meyer2020}. Such trials can improve the statistical efficiency compared to separate trials because of the flexible features such as stopping treatments early for futility or efficacy, adding new treatments to be tested during the course of the trial, and sharing control groups   \cite{Saville2016}. One controversy in the analysis of such trials is the use of non-concurrent controls in treatment versus control comparisons.
For an experimental treatment that enters the platform trial at a later time point, we denote control patients who were recruited before the experimental treatment entered the platform as {\em non-concurrent controls} and patients who are recruited to the control when the experimental treatment is part of the platform as {\em concurrent controls} for the specific treatment arm. Unadjusted treatment-control comparisons that pool concurrent with non-concurrent controls can be biased if there are time trends in the control data, e.g., due to a change in standard of care, change in the patient population, or other external changes such as seasonal effects or a pandemic \cite{Dodd2021,Lee2021, Collignon2021,ICH_9,ICH_10}.

As two hypothetical trials where non-concurrent controls could be incorporated in the analysis, consider the settings of platform trials in depression and non-alcoholic steatohepatitis with a continuous or binary outcome, respectively.

\paragraph{Platform trial for Major Depressive Disorder}

Major Depressive Disorder (MDD) is a common psychiatric disorder 
and is a leading cause of burden of disease and years lost due to disability worldwide 
\cite{otte2016major}. 
While safe and effective pharmacological treatments for depression are available, they only achieve sufficient symptom reductions in about half of the patients \cite{sforzini2022delphi}. Importantly, the major drug classes in this area were discovered and developed decades ago and there have been very few successful attempts to develop novel drugs since. More recently, however, a number of candidate drugs with different putative mechanisms of action have been evaluated in pre-clinical and early stage clinical studies. Thus, platform trials that allow for assessing multiple medications simultaneously have the potential to substantially speed up drug development in this area 
\cite{gold2022platform}. 
Blinding is particularly critical in this indication 
because of the types of outcomes used (e.g. interview-based ratings and patient-reported outcomes), expectation bias and typically large placebo responses in MDD trials. Consider a platform trial in MDD where experimental treatment arms are compared with a common control. The primary endpoint is the change in a continuous clinical rating scale, such as the Montgomery-Åsberg Depression Rating Scale (MADRS), from baseline to six weeks post randomization. 
To reduce the sample size and the number of patients assigned to the control group, the use of non-concurrent controls could be considered. However, time trends in the placebo response cannot be excluded in this setting. A specific source of such time trends can be expectation bias, which may vary, depending on the allocation probability to control and the number and type of treatments currently in the platform. Especially, if the allocation probability to control is low and treatments that are perceived as very promising enter the platform, the expectation bias may increase. Therefore, adjusting for potential time trends would be essential to obtain robust results.

\paragraph{Platform trial for Non-Alcoholic Steatohepatitis}

Non-Alcoholic Steatohepatitis (NASH) is currently an area of high unmet medical need with no approved therapies in Europe and the United States \cite{fraile2021non}.  To facilitate and accelerate the identification of the most effective and promising novel treatment options for participants with NASH, multiple potential novel therapies and combinations thereof can be tested in platform trials.
As example consider a Phase 2b platform trial where the primary efficacy endpoint is a  binary endpoint indicating whether the patient responded or not to treatment  at 48 weeks. As the determination of response requires paired biopsies at baseline and at the end of the study, there is a high patient burden and limiting the sample size is important, especially from a patients perspective. 
Therefore, the use of non-concurrent controls can be considered in the analyses to reduce the number of patients required in the clinical trial. However, also in this setting time trends can occur, e.g., due to changes in the patient population or standard of care, but also due to variability in the assessment of the endpoint \cite{rowe2022placebo}. 

Lee and Wason \cite{Lee2020} investigated linear regression models to estimate the treatment effect of interest in trials with 
continuous data that include a factor corresponding to time to adjust for potential time trends. 
They demonstrated in a simulation study that using regression models that adjust for 
time trends by means of a step-wise function leads to unbiased tests, even if the true time trend is linear rather than step-wise. This holds if the linear regression model is fitted to data of the tested experimental treatment group and the control group as well as to the data of  all other trial arms. However, only if the model is fit based on data of all trial arms, the inclusion of non-concurrent controls improves the power compared to an analysis using concurrent controls only.

In this article, we derive the conditions under which model-based time trend adjustments control the type 1 error rate and increase the statistical power for treatment-control comparisons.  In particular, we consider the simple setting of a two-period platform trial that starts with a single treatment and a control to which another treatment arm is added in the second period.

We consider the regression model fitted with data from all arms and modelling a common step-wise time trend. For this model we show that treatment effect estimates are unbiased {\textit{if the time trends in all arms are equal and additive in the model scale}}. This holds, regardless of the specific functional form of the time trends. For the associated hypothesis tests, we show that for binary endpoints and logistic regression models they asymptotically control the type 1 error rate under the above assumptions. For continuous data analysed with the standard linear model, type 1 error control holds asymptotically, if block randomisation is used.  We show that, under the above conditions, the model-based analysis incorporating non-concurrent controls increases the power of the trial as well as the precision of treatment effect estimates compared to analyses based on concurrent controls only. 

We also investigate the properties of methods when time trends differ across treatment arms or are not additive on the model scale. We show through simulations that heterogeneous time trends across groups can lead to an inflation of the type 1 error rate and biased treatment effect estimates. 
Furthermore, for binary data and logistic regression models, where the validity of the testing and estimation procedures relies on the assumption of equal time trends  on the log odds scale of the regression model, we demonstrate that equal time trends in other scales do not guarantee type 1 error control. 
In addition, we consider alternative regression models, as models allowing for different step-wise time trends between treatments, or regression models fitted with data of the tested treatment arm only. However, they only marginally increase the power of hypothesis tests (only due to larger degrees of freedom for linear models).

Objectives and requirements for confirmatory clinical trials (mainly phase III) and early phase trials are different. This extends to early and late phase platform trials. In spite of these differences, however, the issues we address in this paper are all related to the handling of potential time trends in the accumulating data. We chose to illustrate them using a framework of type I error control, but also investigate the bias in treatment effect estimates. These considerations are of different importance in different phases of drug development, but should not be ignored in any phase.

\section{Methods} \label{sect_methods}

\begin{figure}[h!]  
	\includegraphics[width=0.8\linewidth]{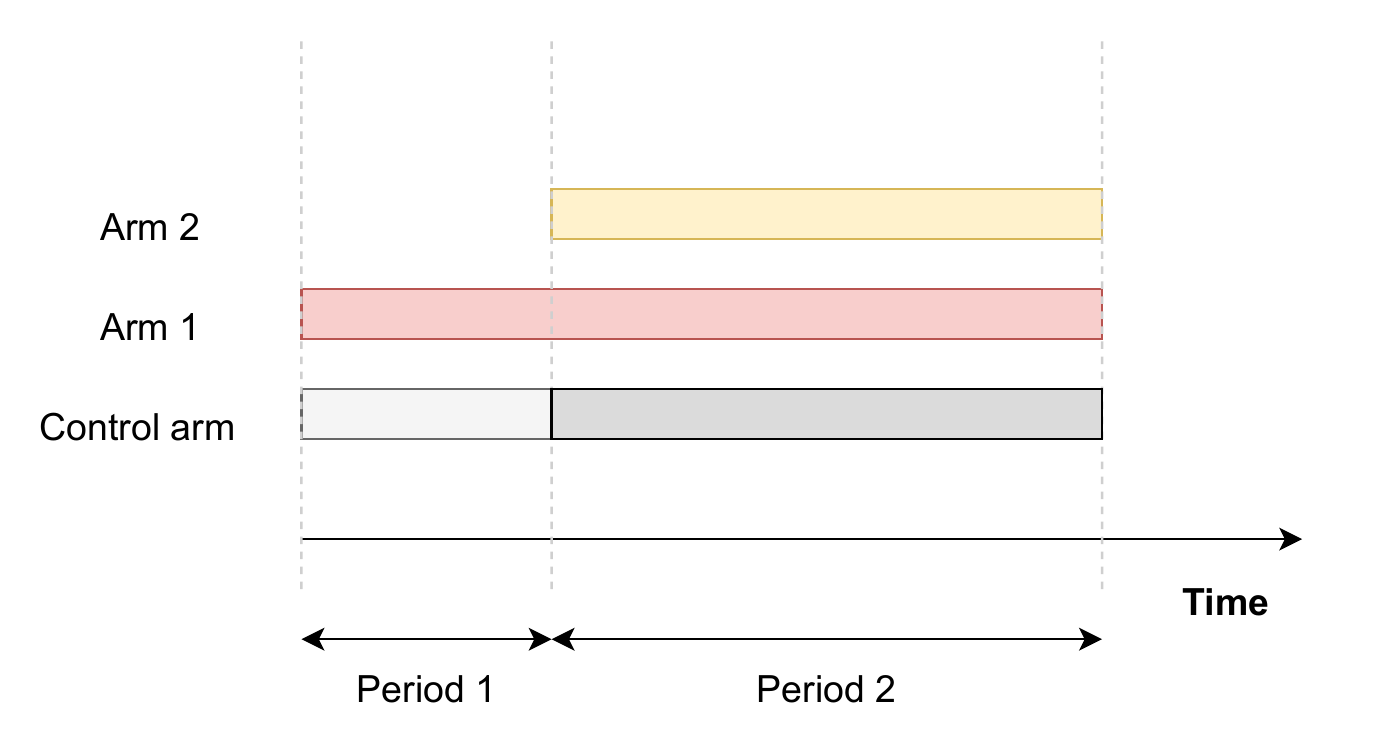}
	\caption{Platform trial scheme. Example platform trial with initially a treatment group (arm 1) and a control group in period 1; and with a new treatment (arm 2) starting in period 2. Light grey represents non-concurrent controls with respect to the new treatment, dark grey represents concurrent controls.}
	\label{fig:trial_scheme}
\end{figure}

To capture the principles of the use of non-concurrent controls, it is sufficient to analyze a simplified model. We therefore consider a randomised, parallel group platform trial, initially comparing a single experimental treatment ($k=1$) to a control ($k=0$), as in \cite{Lee2020}. After $N_1$ patients have been recruited, a new experimental treatment arm ($k=2$) is added to the trial with the intention of comparing this new added arm 2 with the common control. Thus the trial is divided into two periods ($s=1,2$)  before and after the addition of the new arm. See Figure \ref{fig:trial_scheme} for an illustration of the design. 
Furthermore, let $N_2$ denote the total sample size in period 2  and $N=N_1+N_2$ the overall sample size.

We focus on the hypothesis tests and treatment effect estimates comparing treatment 2 to control. We assess the impact of time trends on the validity of inference when incorporating non-concurrent controls according to different adjustment methods. We estimate and test the treatment effect of treatment 2 compared to control based on regression models where the time trend is modelled as a step function.   We consider the regression model investigated in \cite{Lee2020} for continuous data, fitted to data from all treatment arms and the control. In addition, we also investigate a model including an interaction between treatment and time period to allow for a different time trend in treatment arm 1 compared to the other groups. Furthermore, we investigate the extension to binary data. Let $Y_j$ denote the response (binary or continuous) for patient $j$, where $j= 1, ..., N$ is the patient index corresponding to the order in which patients have been enrolled into the trial, and let $k_j$ be the treatment patient $j$ received ($k_j=0,1,2$).

\subsection{Model-based time trend adjustments}

To estimate the treatment effect of treatment 2 against the control,
we fit a regression model, adjusting for time by a step function, based on data from all treatment arms
\begin{align}  \label{ALLTC-Step} 
	g(E(Y_j))&= \eta_0 + \sum_{k=1,2} \theta_k \cdot I(k_j = k) + \nu \cdot I(j>N_1)  
\end{align}
where $\eta_0$ is the response in the control group in the first period, $\theta_{k}$ is the effect of treatment $k$  compared to control, and $\nu$ denotes a step-wise time effect between periods 1 and 2. 
For continuous data, $g(\cdot)$ is the identity function and  we fit a standard linear model as specified in (\ref{ALLTC-Step}) and perform the corresponding t-test to test the one-sided null hypothesis $H_{02}:\theta_2\leq 0$ (assuming that larger values correspond to better outcomes). For binary data, $g(\cdot)$ is the logit link function and $H_{02}$ is tested based on a logistic regression. Also note that the step-wise function, $I(j>N_1)$, amounts to a stratified model with time periods as strata and a common treatment effect across strata.

Model \eqref{ALLTC-Step} fits a common time trend across all treatment arms. To relax this assumption, we consider an additional model that includes an interaction effect, allowing for a different effect of time in treatment arm 1 compared to control and treatment 2. The resulting model  
is given by
\begin{align} \label{ALLTCI-Step} 
	g(E(Y_j))&= \eta_0 + \sum_{k=1,2} \theta_k \cdot I(k_j = k) + \nu \cdot I(j>N_1) 
	+ \eta \cdot  I(k_j = 1)\cdot I(j>N_1)  
\end{align}   

Because a common time trend in the control and treatment 2 is modelled, also in this model the treatment effect of treatment 2 is given by $\theta_2$. 
Also note that in the above models the effect of time on the response is assumed to be additive for continuous data, and multiplicative on the odds ratio scale for binary data.

\subsection{Tests based on concurrent controls only and based on naively pooling controls}

For comparative purposes, we additionally consider the pooled test that does not account for time trends and the test based on concurrent controls only comparing treatment 2 to control using a t-test for continuous and a logistic regression (with factor treatment) for binary data.
We apply these tests using data of treatment 2 and pooling concurrent and non-concurrent controls (pooled analysis) or using concurrent control data only (separate analysis).

\section{Results}  

In this section we report results on the properties of the model based estimation and testing procedures based on model (\ref{ALLTC-Step}). In Section \ref{sssec:estmodperiod} we rewrite the treatment effect estimator for treatment 2 of the linear model as a weighted sum of period-wise per-group means. Based on this representation, we derive the reduction in variance of this estimator compared to the estimator based on concurrent controls only. 

In Section  \ref{sssec:testmodperiod} we discuss further properties of the model-based estimators and tests under the assumption that the time trends are equal across treatment groups and additive on the model scale. We show that the model based treatment effect estimator is unbiased, even if the shape of the time effect is mis-specified and deviates from the step function. In addition, we give conditions on the randomisation or testing procedure, that guarantee control of the type 1 error rate of the model based hypothesis test. 

Finally in Section \ref{sssec:sim} we illustrate the procedures in a simulation study and quantify the potential inflation of type 1 error in the model based hypothesis tests
if time trends are not equal across treatment groups or not additive on the model scale.

\subsection{Testing and estimation in model-based approaches}  

\subsubsection{Estimation under model-based period-wise adjustments\label{sssec:estmodperiod}}

To illustrate how the models \eqref{ALLTC-Step} and \eqref{ALLTCI-Step} estimate the treatment effect of treatment 2 compared to the control when adjusting for potential time trends, we write the regression model estimators as a weighted sum of the period-wise per-group estimates \cite{Lee2020,elm2012flexible}. 
Let $n_{k,s}$ and $\bar{y}_{k,s}$ be the sample sizes and sample means of the observations per arm and period ($k=0,1,2$ and $s=1,2$). Thus, in our setting, $ n_{k,s} \ge 1 $ except for $ n_{2,1} = 0$, and $N_s=\sum_{k=0,1,2} n_{k,s}$.
The estimate of $\theta_2$ is given by
$$
\hat\theta_2 =\sum_{k=0,1,2, s=1,2}w_{k,s}\bar{y}_{k,s},
$$
with weights $w_{k,s}$ defined below. 
The derivation of the weights can be found in the supplementary material (Section A).

First note that for model \eqref{ALLTCI-Step} based on all data and including an interaction effect, 
the matrix $w_{k,s}$ is given by
$$\begin{array}{r|rr}
k\backslash s&1&2 \\\hline
0 & 0 & -1 \\
1 & 0 & 0 \\
2 & 0 & 1 \\
\end{array}$$
so that $\hat\theta_2 = \bar{y}_{2,2} - \bar{y}_{0,2}$, and therefore the non-concurrent controls do not contribute to the treatment effect estimator. 
Hence, we do not consider this model further in this Section.

For model \eqref{ALLTC-Step},
based on data from all treatment arms and adjusting for time by a step function, 
the matrix of weights $w_{k,s}$ is given by

\begin{tabular}{rl}
	\parbox{0.4\textwidth}{
		$$
		\begin{array}{r|rr}
		k\backslash s&1&2 \\\hline
		0 & - \varrho & \varrho - 1 \\
		1 & \varrho & - \varrho \\
		2 & 0 & 1 \\
		\end{array}
		$$
	}
	&
	\parbox{0.6\textwidth}{ 
		where  \
		$\varrho = \frac{ \frac{ 1 }{ n_{0,2} } }{ \frac{ 1 }{ n_{0,1} } + \frac{ 1 }{ n_{0,2} } + \frac{ 1 }{ n_{1,1} } + \frac{ 1 }{ n_{1,2} } }$
	}
\end{tabular}

and thus 
\begin{align*}
	\tilde\theta_2 &= 
	(\bar{y}_{2,2} - \bar{y}_{0,2} ) + \varrho \left[ ( \bar{y}_{1,1} - \bar{y}_{0,1} ) - ( \bar{y}_{1,2} - \bar{y}_{0,2} ) \right]\\
	&=
	\bar{y}_{2,2} - \left\{ 
	(1- \varrho)  \bar{y}_{0,2} + \varrho \left[  \bar{y}_{0,1} + (\bar{y}_{1,2}  -  \bar{y}_{1,1})\right]\right\}\,.    
\end{align*}
Thus, the treatment effect is estimated by the difference in the mean of treatment group 2 and a model-based estimate of the control response in period 2, given by
\begin{equation}\label{estimate_cr}
	\tilde y_{0,2} = (1- \varrho)  \bar{y}_{0,2} + \varrho \left[  \bar{y}_{0,1} + (\bar{y}_{1,2}  -  \bar{y}_{1,1}) \right].
\end{equation} 
Note that this estimate is a weighted average of the mean control response in period 2  and the mean control response in period 1, adjusted by the time trend estimated from treatment 1. The weight of the non-concurrent controls is given by $\varrho$.   
The relative reduction in variance of this estimate compared to the estimator based on concurrent controls only is given by 
\begin{equation*}  
	1-\frac{\text{Var}(\tilde y_{0,2})} {\text{Var}(\bar{y}_{0,2})}=\varrho.
\end{equation*}
Thus, reduction in variance is increasing in the number of non-concurrent control patients, in the number of concurrent patients on arm 1, and in the number of non-concurrent patients on arm 1. In particular, keeping the number of concurrent controls fixed and assuming that $ n_{0,1} \to \infty $, $ n_{1,1} \to \infty $, and $ n_{1,2} \to \infty $, we have  $ \varrho \to 1 $. On the other hand, increasing the number of concurrent control patients but keeping the other sample sizes fixed we have $ \varrho \to 0 $ as $ n_{0,2} \to \infty $, and thus asymptotically no reduction takes place.

Consider now a specific example, with equal randomisation for arms 0 and 1 in each period, that is, $n_{0,1}=n_{1,1}$ and $n_{0,2}=n_{1,2}$, but possibly different for arm 2, so that $ n_{2,2} \ge 1 $ is arbitrary. Then, the reduction in variance is $ \varrho = \frac{1}{2}\cdot\frac{n_{0,1}}{n_{0,1}+n_{0,2}}$ and thus proportional to  the proportion of non-concurrent controls over the total number of controls. Especially, the reduction is delimited above by $ 1/2 $ (as $ n_{0,1} \to \infty $). 
For instance, in the example $n_{0,1}=n_{0,2}=n_{1,1}=n_{1,2}=n_{2,2}/2$ considered by \cite{Lee2020} the weights and the treatment effect estimator are given by

\begin{tabular}{l  l}
	\parbox{0.25\textwidth}{ 
		$$\begin{array}{r|rr}
		k\backslash s&1&2 \\\hline
		0 & -0.25 & -0.75 \\
		1 & 0.25 & -0.25 \\
		2 & 0 & 1 \\
		\end{array}$$
	}
	&
	\parbox{0.7\textwidth}{ 
		, and  \
		$\tilde\theta_2 = ( \bar{y}_{2,2} - \bar{y}_{0,2} ) + 0.25 \left[ ( \bar{y}_{1,1} - \bar{y}_{0,1} ) - ( \bar{y}_{1,2} - \bar{y}_{0,2} ) \right]$.
	}				
\end{tabular}
The model based estimate of the period 2 control response has $25\%$ lower variance than the estimator based on concurrent controls only. 
An example of how weights are defined in the case of equal randomisation also for arm 2, that is, $n_{0,2}=n_{1,2}=n_{2,2}$, can be found in the supplementary material.

So that what we are actually estimating when we use models in \eqref{ALLTC-Step} and \eqref{ALLTCI-Step} is a weighted average of the treatment effect of treatment arm 2 over the control in period 2 and the treatment-time interactions between periods.

\subsubsection{Analytical results on the properties of estimators and hypothesis tests\label{sssec:testmodperiod}}

We derive the properties of tests and estimators of the linear and logistic regression model (\ref{ALLTC-Step}) under the assumption that time trends are equal across groups and additive on the model scale such that the data are generated according to the model
\begin{equation} \label{model_datag} 
	g\left(  E(Y_j)\right) = \eta_0 + \sum_{k=1,2} \theta_k \cdot I(k_j = k) + f(t_{j}),
\end{equation}
where  $Y_j$, $g()$, $\eta_0$ and $\theta_{k_j}$ refer to the  continuous or binary response, the link function (identity and logit functions for continuous or binary responses, respectively), the control response and treatment effects, respectively, as in \eqref{ALLTC-Step} and \eqref{ALLTCI-Step}. For continuous data, we furthermore assume that the error terms in the responses are identically and independently normally distributed with zero-mean and equal variances. 
The term $f(\cdot)$ represents the time trend function and $t_j$ is the calendar time when patient $j$ is enrolled in the trial. 
Note that when $f(\cdot)$ is a step function with a step at the end of period 1, then model \eqref{ALLTC-Step} is correct. Otherwise, \eqref{ALLTC-Step} is a misspecified model because the functional form of the time trend is not correctly modelled. 

For continuous data, the model estimate $\hat\theta_2$ is an unbiased estimate for $\theta_2$, even under mis-specification of the time trend pattern, i.e., if $f(\cdot)$ is not a step function  (see  Section C in the supplementary material  
for a proof). Moreover, as shown in section \ref{sssec:estmodperiod}, the model-based estimate based on all data reduces the variance of the treatment effect estimator as compared to separate analysis, and therefore leads to a gain in power. But, by inspection of \eqref{estimate_cr} it is clear that if the time trend in treatment 1 differs from the trend in the control, the estimate will be biased.

As regards testing, for continuous endpoints and given the functional form of the time trend pattern is correctly specified,  we first notice that the weight $\varrho$ in \eqref{estimate_cr} is chosen such that variance of $\hat\theta_2$ is minimised.  
However, if the time trend pattern is mis-specified
and simple randomisation or random allocation is used, the residual variances of 
the fitted model  \eqref{ALLTC-Step} will depend on the specific shape of the time trend  in the data generating model \eqref{model_datag}. Especially, the residual variance may not be constant over time such that the overall variances in the two periods   
may differ. As the linear model assumes homoscedaticity, this may lead to an underestimation of the standard error of $\hat\theta_2$ and therefore to a type 1 error rate inflation of the corresponding test. See Section F.1 in the supplementary material for an example, where the type 1 error rate is inflated. 

In order to obtain unbiased testing procedures for trials with continuous endpoints, we propose two approaches. 
The first consists of choosing a randomisation procedure that controls the variability over time. We can achieve this by stratifying for calendar time in the randomisation process by means of block randomisation. 
Then, since patients randomised in the same block are enrolled approximately at the same calendar time, asymptotically, the time trend does not introduce additional variation in the estimate. Block randomisation is an equivalent of stratified randomisation, where one stratifies for an important risk factor. In this case, block randomisation coincides with stratifying randomisation by time intervals. However, it is well known that stratified randomisation followed by an analysis that does not adjust for the stratification factor, leads to an overestimation of the variance \cite{matts1988properties}. Thus, the resulting tests will be conservative. If the time trends are very strong, this may result in overly conservative tests.  An alternative approach to this one is to allow for different variances between periods and estimate them separately. 

On the other hand, when using binary endpoints, the variance estimators are consistent even under mis-specification of the trend pattern, $f(\cdot)$. So,  in this case, the treatment effect estimators are asymptotically unbiased and so are the hypothesis tests.

\subsection{Evaluation of methods through simulations \label{sssec:sim}} 

We conducted a simulation study to quantify the gains in efficiency and potential biases of the considered methods under a wide range of scenarios. We investigated settings where model assumptions are met, as well as settings  with a mis-specified model where the time trend does not have a step-wise shape or the time trends differ between treatment arms. Especially, we investigated the performance of the aforementioned analysis approaches  with respect to the type 1 error rate and statistical power for the test of $H_{02}: \theta_2=0$, as well as the bias and mean squared error of the estimator of the effect of treatment 2, $\theta_2$. 

\subsubsection{Design} \label{sect_simstudy} 

We simulated data of a platform trial as described in Section 2, considering the data generating model \eqref{model_datag}.  
We consider that patient index corresponds to the order in which patients are enrolled and that at each unit of time one and only one patient enters  in the trial, so that, $t_j=j$ for $j=1, ..., N$. We assume equal sample sizes per arms and allocations to arms 1:1 and 2:1:1 in periods 1 and 2, respectively. Furthermore, patients were assigned to arms following block randomisation, as proposed in Section \ref{sssec:testmodperiod}. In the first period, we used a block size of 4, while, in the second, a block size of 12. 
We consider three patterns for the time trend function: 
\begin{itemize}
	\item a linear time trend 
	$f(j) = \lambda_{k_j}\frac{(j-1)}{(N-1)}$;
	\item a step-wise time trend 
	$f(j) = \lambda_{k_j} I(j> N_1)$;
	\item and an inverse-U time trend: 
	$f(j) = \lambda_{k_j}\frac{(j-1)}{(N-1)}\left(I(j\leq N_{p})- I(j> N_{p})\right)$; 
\end{itemize}
where $\lambda_{k_j}$ quantifies the strength of the time trend in arm $k_j$ and $I(\cdot)$ the indicator function. 
Note that if $\lambda_{k_j}$ takes different values depending on the treatment arm, then we have different time trends between arms.
For the inverse-U time trend, $N_{p}$ denotes the point at which the trend turns from positive to negative. We consider three different values for this point, it  can be either when half of the patients in period 1 have been recruited ($N_p = N_1/2$), 
after the patients were recruited in period 1 and before the start of the second period ($N_p = N_1$),
or when half of the patients in period 2 have been recruited ($N_p = N_1 + N_2/2$). In the main manuscript, we show the results for the scenarios with $N_p = N_1+N_2/2$. The results for $N_p = N_1/2, N_1$ can be found in the supplementary material. Note that negative $\lambda_{k_j}$ correspond to a U-shaped (instead of inverse-U shaped) trend.  Figure \ref{fig:cont_response_main} illustrates the changes of means according to these patterns. In the  supplementary material  we also show that considering random instead of fixed entry times $t_j$ does not noticeably change the results.

For treatment 1 and the control group the sample size per group  was set to 125 in each period. The sample size for treatment 2 in period 2 is 250 (such that its sample size matches the overall sample size of treatment 1).
For continuous endpoints, we assumed normally distributed residuals with equal variances $\sigma^2=1$ across treatment arms, zero mean for control arm ($\eta_0=0$), 
simulated trials under the null hypothesis, $\theta_2 =0$, as well as the alternative hypothesis $\theta_2 =0.25$. In both cases, we considered an effect of $\theta_1 =0.25$ for treatment 1. For these effect sizes, the pooled t-test has 80\% power at 2.5\% one-sided significance level. 
For binary endpoints, we considered control response rates of $0.7$, non-effect of treatment 2 under the null hypothesis $\theta_2=0$ (in terms of the odds ratio $OR_2=1$) or a positive effect under the alternative hypothesis ($OR_2=1.8$), and scenarios where treatment 1 has either a positive or a negative effect 
(i.e., $\theta_1>0$ and $\theta_1<0$, respectively).
With the aforementioned configuration for these scenarios, the power to detect an odds ratio $OR_2=1.8$ is 80\% assuming a probability $p_0=0.7$ at 2.5\% significance level using the pooled z-test. 
Finally, we simulated the three time trends patterns mentioned before  
assuming equal time trends across all treatment groups ($\lambda_k = \lambda$, for $k=0,1,2$), and equal time trends between  treatment 2 and control arms and different from treatment 1 arm ($\lambda_0 = \lambda_2 \neq \lambda_1$). We simulated 100,000 replicates for each scenario  to obtain estimates of the type 1 error, power and bias of the treatment effect.

\subsubsection{Simulation results}

In settings where the time trends are equal across treatment arms (i.e., scenarios with $\lambda_k=\lambda$ for $k=0,1,2$), all regression models that fit time trends as a step function control the type 1 error rate for the test of $H_{02}$ and lead to unbiased effect estimates in the examples we investigated.  This holds, if the true time trend is a step function as well as if it is linear or U-inverted. For the model without interaction term \eqref{ALLTC-Step} this confirms the theoretical considerations in the above section as well as the simulation results of \cite{Lee2020}. Furthermore, the test based on \eqref{ALLTC-Step} leads to an improved power compared to the separate analysis using only concurrent controls (see Figure S1 in the supplementary material). Similarly, for binary data we note that in settings where the time trends are equal across arms and given by \eqref{model_datag}, the tests and estimators of the corresponding logistic regression model control the error rate and we see an improved power compared to the separate analysis using only concurrent controls. (Figure S7 in the supplementary material.).

\begin{figure}[h!]  
	\includegraphics[width=0.95\linewidth]{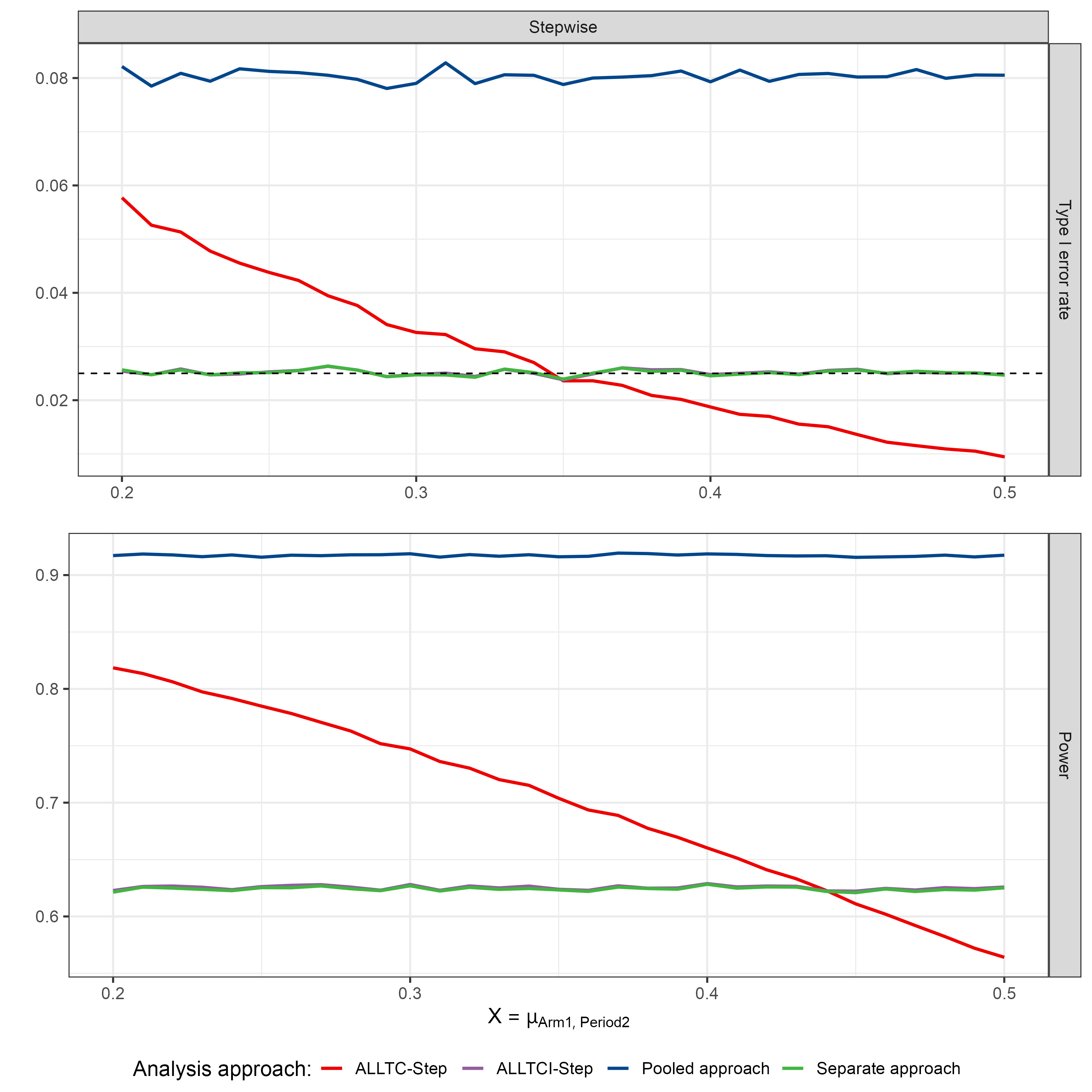}
	\caption{Type 1 error rate and power of rejecting $H_{02}$ for continuous endpoints in the presence of  step-wise trends with respect to the mean in the treatment arm 1 in the second period and according to the model used (see Sect. \ref{sect_methods}) for scenarios presented in Sect. \ref{sect_simstudy}  and described in Table \ref{tab_excont}.
		ALLTC-step refer to models using all treatment data and control and adjusting for time by a step function (see \eqref{ALLTC-Step}), ALLTCI to models using all treatment data and control with interaction between time and treatment arm and adjusting for time by a step function (see \eqref{ALLTCI-Step}), and  pooled and separate approaches refer to t-tests comparing treatment 2 to control using concurrent and non-concurrent control data, and concurrent control data only, respectively. Note that lines for ALLTCI-Step and the separate approach are overlapping.
	}
	\label{fig:cont_step_diff_pow_alpha_main}
\end{figure}

\begin{figure}[h!]  
	\includegraphics[width=0.95\linewidth]{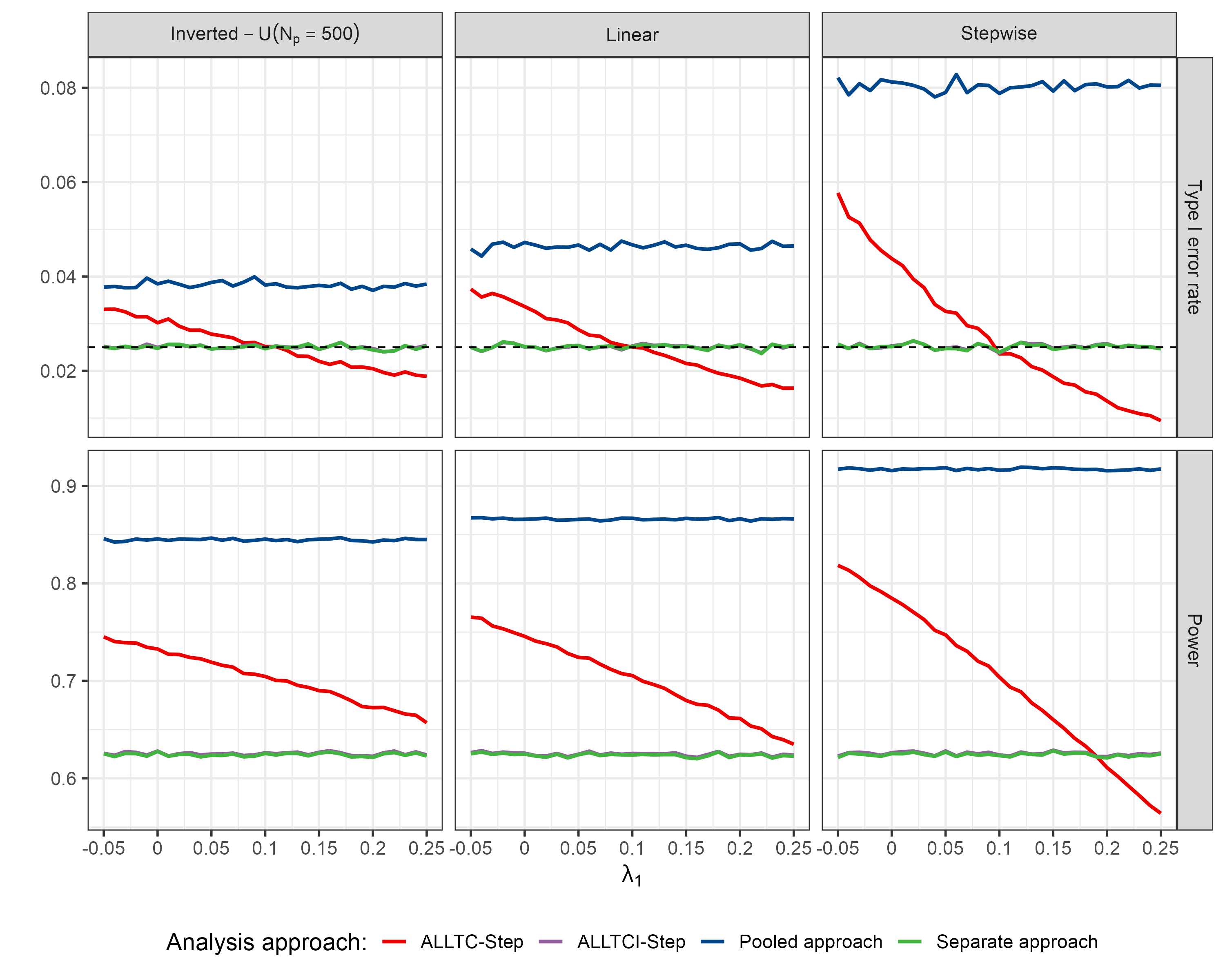}
	\caption{Type 1 error rate and power of rejecting $H_{02}$ for continuous endpoints in the presence of  linear, step-wise and inverted-U time trends (for $\lambda_0 = \lambda_2 = 0.1$) with respect to the strength of the time trend in treatment arm 1 ($\lambda_1$) according to the model used (see Sect. \ref{sect_methods}) for scenarios presented in Sect. \ref{sect_simstudy}. 
		ALLTC-step refers to models using all treatment data and control and adjusting for time by a step function (see \eqref{ALLTC-Step}), ALLTCI refers to models using all treatment data and control with interaction between time and treatment arm and adjusting for time by a step function (see \eqref{ALLTCI-Step}), and  pooled and separate approaches refer to t-tests comparing treatment 2 to control using concurrent and non-concurrent control data, and concurrent control data only, respectively.  
		Lines for ALLTCI-Step and the separate approach are overlapping.}
	\label{fig:cont_all_diff_pow_alpha_main}
\end{figure}

In contrast, if the time trend in treatment 1 differs from the trend in the control arm, control of the type 1 error is no longer guaranteed. For example, suppose a setting as in Table \ref{tab_excont} where the mean outcome in the control arm increases by 0.1 from period 1 to period 2. Then, the type 1 error rate for the test of $H_{02}$ depends on the increase in the mean outcome of treatment 1. Let us consider different values of the mean outcome of arm 1 in period 2 as in Figure \ref{fig:cont_step_diff_pow_alpha_main}.
If the increase in the mean outcome of treatment arm 1 from period 1 to period 2 is lower than in the control, the type 1 error rate may be substantially inflated, reaching type 1 error rates larger than $0.05\%$. If it is larger than 0.1 the procedure becomes strictly conservative, obtaining type 1 error rates below $0.01$, and the power is adversely affected (see Figure \ref{fig:cont_step_diff_pow_alpha_main}). Indeed, the treatment effect estimates will be positively or negatively biased depending on the time trend in treatment 1. Furthermore, if instead of step-wise time trends, the time trends' patterns are  linear or inversed-U, the same behaviour is observed, as is shown in  Figure \ref{fig:cont_all_diff_pow_alpha_main} when varying the strength of the time trend in arm 1 and fix the time trends of control and arm 2.

\begin{figure}[h!]  
	\includegraphics[width=0.95\linewidth]{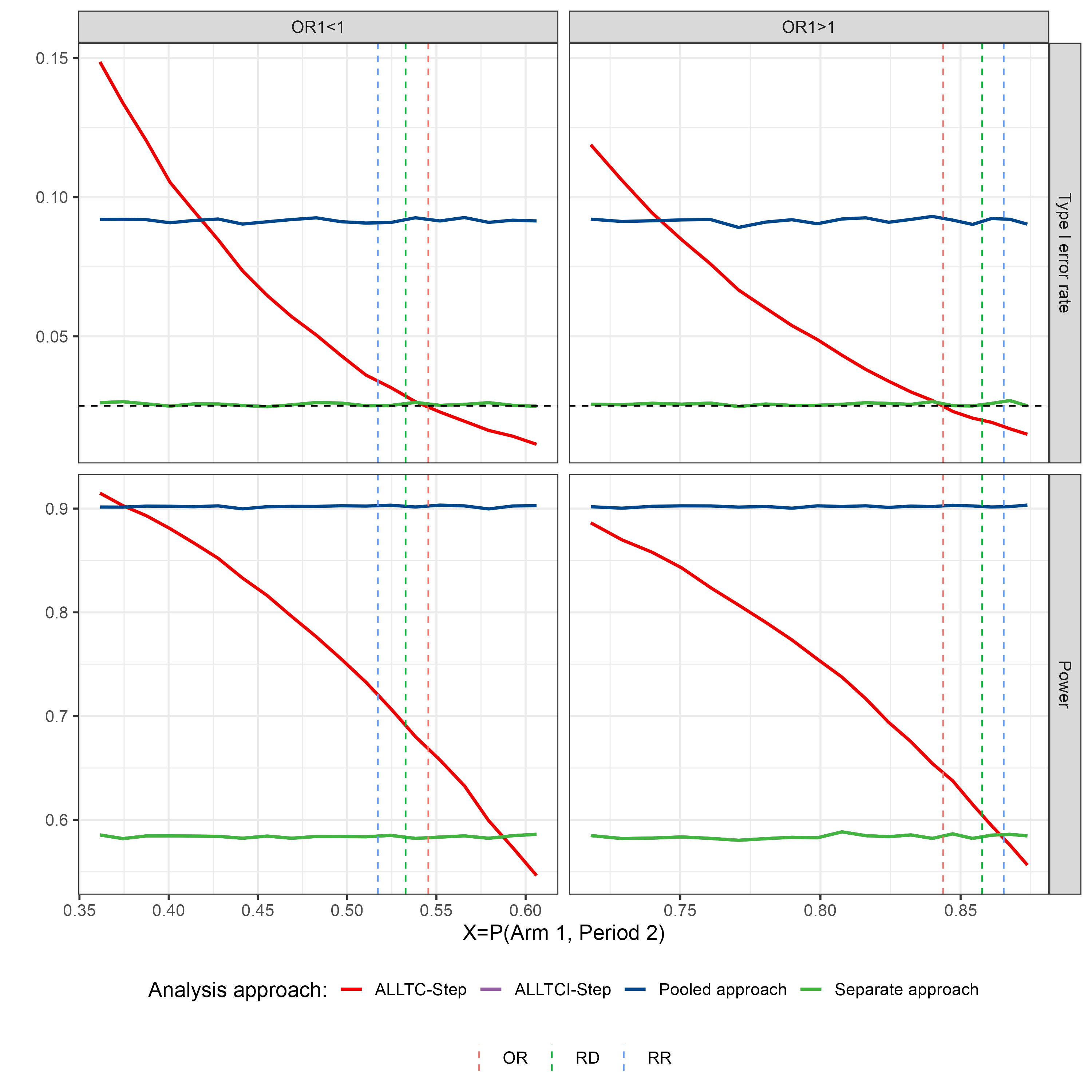}
	\caption{Type 1 error rate and power of rejecting $H_{02}$ for binary endpoints in the presence of  step-wise trends (for $\lambda_0 = \lambda_2$ and varying $\lambda_1$) with respect to the response rate in the treatment arm 1 in the second period and according to the model used (see Sect. \ref{sect_methods})  for scenarios presented in Sect. \ref{sect_simstudy} and described in Tables \ref{tab_exbin1} and \ref{tab_exbin2}. Vertical lines refer to those situations when time trends are equal across arms on odds ratio (OR), risk difference (RD), and relative risk (RR) scales.
		ALLTC-step refers to models using all treatment data and control and adjusting for time by a step function (see \eqref{ALLTC-Step}), ALLTCI refers to models using all treatment data and control with interaction between time and treatment arm and adjusting for time by a step function (see \eqref{ALLTCI-Step}), and  pooled and separate approaches refer to logistic regression models without adjusting for time comparing treatment 2 to control using concurrent and non-concurrent control data, and concurrent control data only, respectively. 
		Lines for ALLTCI-Step and the separate approach are overlapping.}
	\label{fig:bin_step_diff_pos_pow_alpha_main}
\end{figure}

A similar result can be found for binary data when considering a fixed increment in the response rate from period 1 to period 2 of the control arm and varying the response rate of the treatment arm 1 assuming to have a beneficial effect (see Table \ref{tab_exbin1}). 
If the time trend in treatment 1 is smaller than in the control arm on a multiplicative odds ratio scale the  type 1 error rate may be inflated. If the time trend is larger, the test becomes strictly conservative (and loses power) (see second column in Figure \ref{fig:bin_step_diff_pos_pow_alpha_main}). 
Note that this implies that - given for treatment 1 the alternative holds - when time trends are equal between control and treatment 1 on a risk ratio or risk difference scale (vertical lines in blue and green colors in Figure \ref{fig:bin_step_diff_pos_pow_alpha_main}, respectively), the type 1 error of the test for treatment 2 is not maintained.

\begin{figure}[h!]  
	\includegraphics[width=0.95\linewidth]{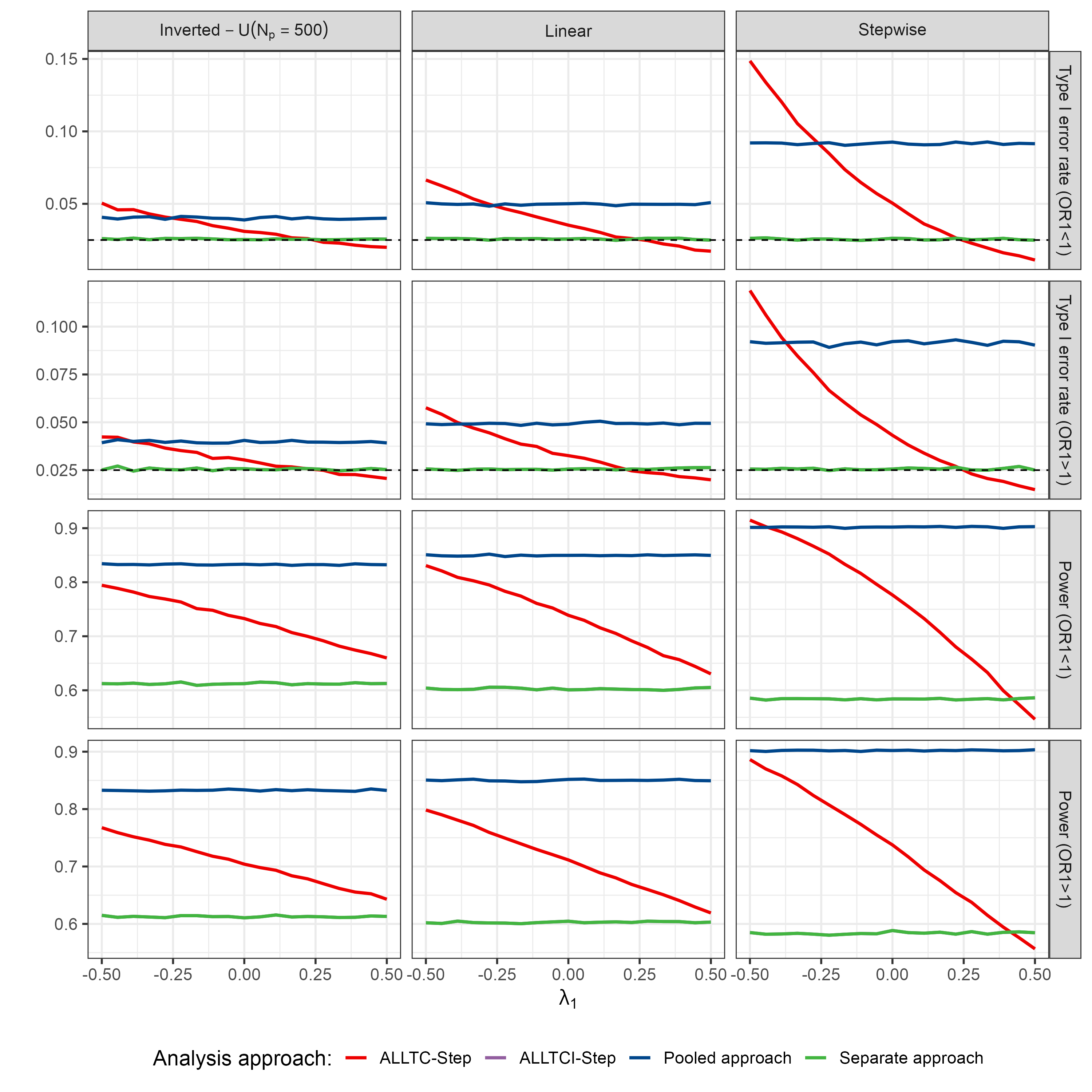}
	\caption{Type 1 error rate and power of rejecting $H_{02}$ for binary endpoints in the presence of  linear, step-wise and inverted-U time trends (for $\lambda_0 = \lambda_2$) with respect to the strength of the time trend in treatment arm 1 ($\lambda_1$) 
		and according to the model used (see Sect. \ref{sect_methods})  for scenarios presented in Sect. \ref{sect_simstudy}. 
		ALLTC-step refers to models using all treatment data and control and adjusting for time by a step function (see \eqref{ALLTC-Step}), ALLTCI refers to models using all treatment data and control with interaction between time and treatment arm and adjusting for time by a step function (see \eqref{ALLTCI-Step}), and  pooled and separate approaches refer to logistic regression models without adjusting for time comparing treatment 2 to control using concurrent and non-concurrent control data, and concurrent control data only, respectively.
		Lines for ALLTCI-Step and the separate approach are overlapping.}
	\label{fig:bin_all_diff_pos_pow_alpha_main}
\end{figure}

If we consider the same situation as above, but instead of having a beneficial effect in treatment 1, this shows a negative effect as compared to the control (see Table \ref{tab_exbin2}), the same behaviour is observed but inflation is much higher, even reaching levels above $0.10$ (see first column in Figure \ref{fig:bin_step_diff_pos_pow_alpha_main}). 
Moreover, similar results are observed when trends are linear or inversed-U rather than piece-wise as shown in Figure \ref{fig:bin_all_diff_pos_pow_alpha_main}. Note, however, that since in our simulations the strength of the trend ($\lambda_1$)  takes the same value in all three patterns, it has a greater impact on the piece-wise trend, 
since changes are more gradual in the two cases with continuous time trends. 

An option to address the type 1 error inflation is to apply the extended model \eqref{ALLTCI-Step} including an interaction term between period and treatment 1. Indeed, as shown in Figures \ref{fig:cont_all_diff_pow_alpha_main} and \ref{fig:bin_all_diff_pos_pow_alpha_main}, for the above example the extended model controls the type 1 error rate, both for continuous as well as binary data. However, applying this model does not lead to a relevant improvement in power compared to the analysis based on concurrent data only.
Indeed for both, the linear and the logistic regression models, data of non-concurrent controls do not contribute to the treatment effect estimate. For the linear model, however, they contribute to the variance estimate. 

Note that in this simulation study we focused on how time trends affect the operating characteristics when using model-based adjustments. However, also other deviations from the model assumptions not considered in the simulation study (as, e.g., heteroscedasticity) can affect the type I error rate. Especially, for unequal sample sizes in the control and treatment groups and unequal variances between groups or periods, the pooled variance estimator of the linear model is biased. Depending on the scenario, this can lead to  an increased type 1 error rate but also strictly conservative testing procedures. To account for heterogeneous variances, a  model allowing for different variances across periods and treatment groups can be fitted.

\subsubsection{Additional simulation results for trials with three periods} 

Above we considered platform trials starting with a single treatment and a control, where a second treatment arm enters later but ends at the same time as treatment arm 1. Consequently, the factor period in the regression model for this platform trial has only two levels.  

In addition, we performed a simulation study for trials where recruitment to arm 1 ends before the end of the platform trial. Then, the trial is divided into three time periods, where the first and last only involve one treatment and the control but in the second participants are randomised to both treatment arms and the control (see Figure S16 in the supplementary material).
Overall, we find that the properties described for the two period trial design also apply to the scenarios with three periods considered in the simulations. In particular, under equal time trends, the type 1 error is controlled when modeling time trends with step functions based on data from all arms using  a variant of model \eqref{ALLTC-Step} where period has three levels. If the time trend in treatment 1 differs from that in treatment arm 2 and the control, the type 1 error rate may be inflated. However, in the considered examples the type 1 error rate inflation is less pronounced than in the two-period design because the overlap between arms 1 and 2 is smaller. As a consequence, the non-concurrent controls have less weight in the test statistics computed from the step function model. However, this implies that the power gain compared to an analysis based on concurrent controls only is smaller than in the design with two periods (see the supplementary material for further details).

\section{Further modeling approaches} \label{sect_ftmethods} 

In previous sections, we modeled time trends with step functions, based on data from all arms. In this section, we introduce some alternative methods and investigate their properties. For the simulation results for these methods see the supplementary material  (see Figures S4 and S12). 

If rather than using data from all treatment arms as in \eqref{ALLTC-Step} and \eqref{ALLTCI-Step}, one wants to compare the efficacy of treatment 2 against control using data from the control arm in both periods and data of treatment 2  (which is collected in the second period, only),  a regression model fitted by the observations $\left\{Y_j, j \in \{1,\ldots,N\}| k_j=0,2\right\}$ can be used. 
This regression model 
is given by 
\begin{align}    \label{TC-Step}
	g(E(Y_j)) &= \eta_0 +   \theta_2 \cdot I(k_j = 2) +  \nu \cdot  I(j>N_1) 
\end{align}    
where the period effect is fitted in the model in a step-wise way as before.

As noted by Lee and Wason \cite{Lee2020}, linear models that only use data of treatment 2 and control arms (as in \eqref{TC-Step}) maintain the type 1 error rate and lead to unbiased estimators of the treatment effect.  This holds, as long as the variance of the residuals is the same in the two periods. Furthermore, type 1 error rate control is maintained, even if there is a different time trend in treatment arm 1. However, there is no relevant gain in power as compared to the separate analysis. This holds, because the non-concurrent control data do not contribute to the treatment effect estimate.
For the linear model, however, the non-concurrent control data contribute to the variance estimator. Though, only for very small sample sizes, this has an impact on the power. Therefore, the incorporation of the non-concurrent control by means of this approach does not increase the trial's efficiency.  

Instead of assuming a step-wise effect of time over periods, models that linearly adjust for time can be used.
Here we assume that patients arrive at a uniform pace, and patient index is proportional to time. Versions of the regression models \eqref{ALLTC-Step}, \eqref{ALLTCI-Step} and \eqref{TC-Step} 
with linear trends for time are obtained by replacing the step function term in the models above by a linear effect function, and are as follows:
\begin{align}  \label{ALLTC-Linear} 
	g(E(Y_j)) &= \eta_0 + \sum_{k=1,2} \theta_k \cdot I(k_j = k) + \gamma \cdot j  \\
	g(E(Y_j)) &= \eta_0 + \sum_{k=1,2} \theta_k \cdot I(k_j = k) + \gamma \cdot j   +   \eta \cdot j \cdot I(k_j = 1)  \label{ALLTCI-Linear} \\ 
	g(E(Y_j))& = \eta_0 +   \theta_2 \cdot I(k_j = 2) + \gamma \cdot j \label{TC-Linear} 
\end{align}   

Similarly to what was previously noted, models incorporating data from all treatment arms (\eqref{ALLTC-Step} and \eqref{ALLTC-Linear}) control the type 1 error rate and can substantially improve the power under the assumption of equal time trends in all treatment arms in the scale of the model. However, these models in \eqref{ALLTC-Linear} additionally rely on the correct specification of the time trend in the model. If correctly specified, that is, when time trends are linear,  such models achieve higher power as compared to those modeling time as a step function (as in \eqref{ALLTC-Step}) while controlling the type 1 error rate. But, error inflation occurs when the time trend is non-linear. Furthermore, when time trends differ between arms, the inflation is even higher than when time is modeled as a step function. 
In the same way,  models \eqref{ALLTCI-Linear} and \eqref{TC-Linear} also yield biased results and type 1 error inflation when the parameterisation of the time in the model is incorrectly specified. 

\section{Discussion} \label{sect_discussion}

We investigated frequentist, model-based approaches to adjust for time trends in platform trials utilizing non-concurrent controls. We examined conditions under which the model-based approaches can successfully adjust for time trends for the simple case of a two-period trial with two experimental treatments and a shared control. The model-based approach including data from all arms and fitting a common time trend as a step function, controls the type 1 error rate and improves the statistical power in all considered scenarios where the time trend is equal across arms and additive in the model scale. 
This holds asymptotically, even if the time trend is not a step function but has a different shape and if block randomisation is used. Suppose, in contrast, random allocation is used, and the time trends are not step functions. In that case, an inflation of the type 1 error in the linear model can arise because the variance of the residuals may differ across periods (but only a single variance term is estimated by the model). A simple fix for this is to fit a model that allows different residual variances per period. 
On the other hand, type 1 error control is lost for the above step function models if the time trends are different between arms or not additive in the scale of the model. The amount of potential type 1 error rate inflation depends on the size of the differences between the time trends across arms. If the differences are sufficiently small, the type 1 error rate is not substantially affected (see Figure \ref{fig:cont_step_diff_pow_alpha_main}).  

Furthermore, we considered models, modelling time trends as linear functions or as step functions with interaction terms for treatment arm and period. In addition, we investigated a step function model fitted on the control group data and data of treatment arm 2  only. We show that these models do not lead to a noticeable gain in power or are less robust with respect to the control of the type 1 error rate. 

For the assessment of estimators in clinical trials, the estimand framework \cite{ICH_9R1} has become an important tool. While the estimand, defined as the target of estimation, is derived from the trial objective and is therefore not directly affected by the use of non-concurrent controls, the properties of  estimators depend on if and how non-concurrent controls are employed in the estimation of treatment effects \cite{collignon2022estimands}. Time trends can affect estimators through several aspects of the data generating process in a clinical trial, such as the study population, the measurement of endpoints, or the impact of intercurrent events.   

Also, in the hypothetical platform trials in MDD and NASH, considered in the introduction, time trends can occur. The model-based analysis can fully adjust for time trends only if the assumption of homogeneity and additivity of the time trends holds. While one can empirically assess these assumptions based on the observed data, such assessments may have limited sensitivity if trials are not powered for this objective. On the other hand, experience from previous trials and subject matter knowledge may provide justification for the assumption of equal time trends. 
Heterogeneous time trends across treatment arms may occur in settings where several treatments show an effect compared to control, but the external factors causing the time trend do not equally affect the efficacy of all considered treatments. Examples are vaccine trials, where different variants of the targeted virus are predominant over time and vaccine efficacy depends on the variant. Another example is shifts in the population in settings, where treatment efficacy is not homogeneous across the population. In particular, patients recruited early in a trial may already be known by the investigator(s), while newly diagnosed patients may enter later on. In addition, time trends may not be additive in the scale considered in the model, e.g., if there are ceiling effects in the outcome scale (as it happens, for instance, in ADAS-COG in Alzheimer's disease), if the treatment effect is not additive to the placebo effect \cite{boehm2017does} (as it may occur in the example of MDD), or, for binary outcomes, when the time trends are equal on different scales (as the risk ratio instead of the odds ratio scale). However, even if there is a risk of bias, the use of non-concurrent controls can lead to more precise estimates in terms of mean squared error because of a bias-variance trade-off. Especially, in settings with small sample sizes, the reduction in variability may outweigh the potential bias introduced. The extent of time treatment interactions in actual clinical trials will depend on the specific context and it would be important to analyse recent platform trials in this respect to better understand the plausibility of different scenarios. 
The scenarios we considered in the simulation study used thresholds for type 1 and type 2 error control that are frequently used in traditional confirmatory trials. In early phase platform trials different choices would be made reflecting the different trade-offs involved. Furthermore, the trial may not be designed with error rate control as a major focus.


While we considered only a simple scenario, these models can directly be extended to trials with more treatments and periods. In this case, a period defines the time span where the set of treatments in the platform trial does not change. Also in this setting, the assumption of equal time trends that are additive on the model scale is needed to obtain unbiased estimators and conservative testing procedures. If time trends differ, the amount of bias and type 1 error rate inflation will depend on the weight the treatment arms with deviating time trends have in the final test statistics. In addition, if there are different deviating time trends across arms their impact may cancel out, depending on the specific scenario. 

In this paper, we focus on trials with no interim analysis, where the sample size per arm is fixed. However, platform trials often allow for interim analyses with the possibility to stop an arm early, either due to futility or due to an early rejection of the null hypothesis. In trials with such interim analyses, the factor period, defined by the time periods between an entry or exit of an arm, will depend on the interim results. This introduces additional complexity 
as it affects the joint distribution of the independent factor period and the dependent variable in the model and can lead to biased treatment effect estimates and an increased type 1 error rate.

We delimited our investigations to analyses based on linear and logistic regression models. 
To extend the approach to other types of data (e.g., time to event endpoints,  counts and clustered data), the respective appropriate regression models, including time as a step function, could be applied. In addition, to address confounding with measured baseline variables, the regression models can be extended to adjust for other covariates.

Another extension are models that model time trends with more general smooth functions such as splines. Furthermore, non-parametric Bayesian approaches have been proposed to estimate the effect of time, smoothing estimates of time trends by borrowing information between periods \cite{Saville2022}. However, also these methods correctly adjust for time trends only if they are equal in all arms. 

A further approach that has been proposed to adjust for time trends are randomisation tests. For two-armed trials with covariate and response adaptive randomisation, for example, they have been shown to robustly control the type 1 error rate in the presence of time trends \cite{Simon2011}. These tests can be extended to multi-armed trials \cite{wang2020randomization} using conditional randomisation tests, where for each treatment-control comparison, the re-randomisation is limited to the respective treatment and control. However, this conditional re-randomisation is deterministic for the non-concurrent controls in platform trials. Therefore, including non-concurrent controls does not improve the power of the conditional randomisation test compared to a conditional randomisation test based on concurrent controls only. Consequently, this approach would not be beneficial in increasing the efficiency of platform trials by incorporating non-concurrent controls.  

Non-concurrent controls can be considered as the ideal case of historical data, since they are borrowed from the immediate past, randomised, and are part of the same trial infrastructure \cite{burger2021use}. It is an open question whether the large array of dynamic modeling methods proposed to incorporate historical data \cite{Viele2014,Ghadessi2020,Hall2021,Schmidli2014,Schmidli2020} can be adapted to incorporate non-concurrent controls in platform trials.

\section{Conclusions} 

Model-based analyses that make use of data from all treatment arms including non-concurrent controls are a powerful and fairly robust analysis strategy.
If time trends in the different arms have equal additive effects on outcome on the model scale, a model that adjusts for time as step function based on data from all arms controls the type 1 error rate while increasing the power compared to an analysis with concurrent controls only. For binary data and logistic regression models, type 1 error control is maintained, even if the functional form of the common time trend deviates from the step function. For continuous data, this holds if block randomisation is applied or a model allowing for different variances across periods is chosen. 
However, if time trends differ between arms or are not additive to treatment effects in the scale of the model, the type 1 error control may be lost and treatment effect estimates can be biased.

If non-concurrent controls are included in the analysis of platform trials, we recommend the use of step function models to adjust for potential time trends. 
Only in trials where the environment  can be considered ``stable" for the duration of the trial and there is no time trend in the outcomes (e.g., for short trials with objective endpoints and homogeneous patient populations and in diseases with a long-established standard of care and no recent changes in treatment options), adjustment for time trends may not be necessary, and, as in this scenario the distribution of concurrent and non-concurrent controls are identical, we can simply pool concurrent and non-concurrent controls.  In any case, if non-concurrent data are utilised as the primary analysis, also the analysis using only concurrent control data should be presented as sensitivity analysis.

Whether to use non-concurrent controls or not should be a case by case decision based on the expected bias-variance trade-off and subject-matter considerations such as the duration of the trial and potential changes in the standard of care during the trial. Especially in settings with small sample sizes, the efficiency gained by using step function models to incorporate non-concurrent controls can outweigh potential biases. To assess the bias-variance trade-off, the specifics of the trial, scientific plausibility of heterogeneous time trends across treatments, and robustness of results should be carefully considered.

\subsubsection*{Funding}

EU-PEARL (EU Patient-cEntric clinicAl tRial pLatforms) project has received funding from the Innovative Medicines Initiative (IMI) 2 Joint Undertaking (JU) under grant agreement No 853966. This Joint Undertaking receives support from the European Union’s Horizon 2020 research and innovation programme and EFPIA andChildren’s Tumor Foundation, Global Alliance for TB Drug Development non-profit organisation, Spring works Therapeutics Inc. This publication reflects the authors’ views. Neither IMI nor the European Union, EFPIA, or any Associated Partners are responsible for any use that may be made of the information contained herein.

\subsubsection*{Abbreviations}

OR: Odds ratio, 
ALLTC-Step model: \textbf{all} \textbf{t}reatment - \textbf{c}ontrol arms model (see equation \eqref{ALLTC-Step}),
ALLTCI-Step model: \textbf{all} \textbf{t}reatment - \textbf{c}ontrol arms model with \textbf{i}nteraction (see equation \eqref{ALLTCI-Step}).

\subsubsection*{Availability of data and materials}

R code are available on GitHub through the repository: \url{https://github.com/MartaBofillRoig/NCC_timetrends}. This repository contains the main code and R functions to reproduce the results of the simulation study. The datasets generated and/or analysed are also available in the repository.

\subsubsection*{Competing interests}

Carl-Fredrik Burman is an employee and shareholder of AstraZeneca.
Dominic Magirr declares a competing interest as an employee of Novartis Pharma AG. 
Peter Mesenbrink is an employee of Novartis Pharmaceuticals Corporation and owns both restricted and unrestricted shares of stock in Novartis. 
Peter Jacko and Kert Viele are employees of Berry Consultants, a consulting company specializing in the design of Bayesian and adaptive clinical trials. The rest of the authors declare that they have no competing interests regarding the content of this article.

\subsubsection*{Authors' contributions}

MBR and MP conceived and designed the research and jointly contributed to drafting the article. EG, MBR, MP, PJ and KV drafted the methodological parts. MBR and PK wrote the code for the simulations and PK performed the simulations and prepared figures and tables.  All authors contributed to the interpretation of findings, critically reviewed and edited the manuscript. All authors critically appraised the final manuscript.

\subsubsection*{Additional Files}

Supplementary document is available with technical derivations and proofs, and additional results of the simulation study. 

\subsubsection*{Acknowledgements}

The authors thank the reviewers for their constructive comments and suggestions on the manuscript.


%
\bibliography{refs} 




\newpage

\section*{Tables and figures}

\begin{table}[h!] 
	\caption{Example for continuous endpoints. Mean responses for continuous endpoints according to the arm and the period in the presence of  step-wise trends for scenarios presented in Sect. \ref{sect_simstudy} and for $\lambda_0=\lambda_2=0.1$. Here X represents the mean response of arm 1 in period 2, which we vary in Figure \ref{fig:cont_step_diff_pow_alpha_main}.}
	\centering
	\begin{tabular}{|c|c|c|c|c|c|c|}
		\hline
		& \multicolumn{3}{c|}{Period 1} &  \multicolumn{3}{c|}{Period 2} \\
		\hline
		& Control & Arm 1 & Arm 2 & Control & Arm 1 & Arm 2 \\
		\hline
		H0 & 0 & 0.25 & -- & 0.1 & X & 0.1 \\
		\hline
		H1 & 0 & 0.25 & -- & 0.1 & X & 0.35\\
		\hline
	\end{tabular}
	\label{tab_excont}
\end{table}

\begin{table}[h!] 
	\caption{Example for binary endpoints. Responses rates for binary endpoints according to the arm and the period in the presence of  step-wise trends for scenarios presented in Sect. \ref{sect_simstudy} with $\lambda_0 = \lambda_2=0.25$.  Here X represents the observed response rate of arm 1 in period 2, which we vary in Figure \ref{fig:bin_step_diff_pos_pow_alpha_main}. Arm 1 is assumed to have a beneficial effect in period 1, corresponding to $OR_1>1$   ($OR_1 = 1.8$).}
	\centering
	\begin{tabular}{|c|c|c|c|c|c|c|}
		\hline
		& \multicolumn{3}{c|}{Period 1} & \multicolumn{3}{c|}{Period 2} \\
		\hline
		& Control & Arm 1 & Arm 2 & Control & Arm 1 & Arm 2 \\
		\hline
		H0 & 0.7 & 0.81 & -- & 0.75 & X & 0.75 \\
		\hline
		H1 & 0.7 & 0.81 & -- & 0.75 & X & 0.84 \\
		\hline
	\end{tabular}
	\label{tab_exbin1}
\end{table}

\begin{table}[h!] 
	\caption{Example for binary endpoints. Responses rates for binary endpoints according to the arm and the period in the presence of  step-wise trends for scenarios presented in Sect. \ref{sect_simstudy} with $\lambda_0 = \lambda_2=0.25$.  Here X represents the observed response rate of arm 1 in period 2, which we vary in Figure \ref{fig:bin_step_diff_pos_pow_alpha_main}. Arm 1 is assumed to have a negative effect in period 1, corresponding to $OR_1<1$ ($OR_1 = 0.4$).}
	\centering
	\begin{tabular}{|c|c|c|c|c|c|c|}
		\hline
		& \multicolumn{3}{c|}{Period 1} & \multicolumn{3}{c|}{Period 2} \\
		\hline
		& Control & Arm 1 & Arm 2 & Control & Arm 1 & Arm 2 \\
		\hline
		H0 & 0.7 & 0.48 & -- & 0.75 & X & 0.75 \\
		\hline
		H1 & 0.7 & 0.48 & -- & 0.75 & X & 0.84 \\
		\hline
	\end{tabular}
	\label{tab_exbin2}
\end{table}

\begin{figure}[h!]  
	\includegraphics[width=0.95\linewidth]{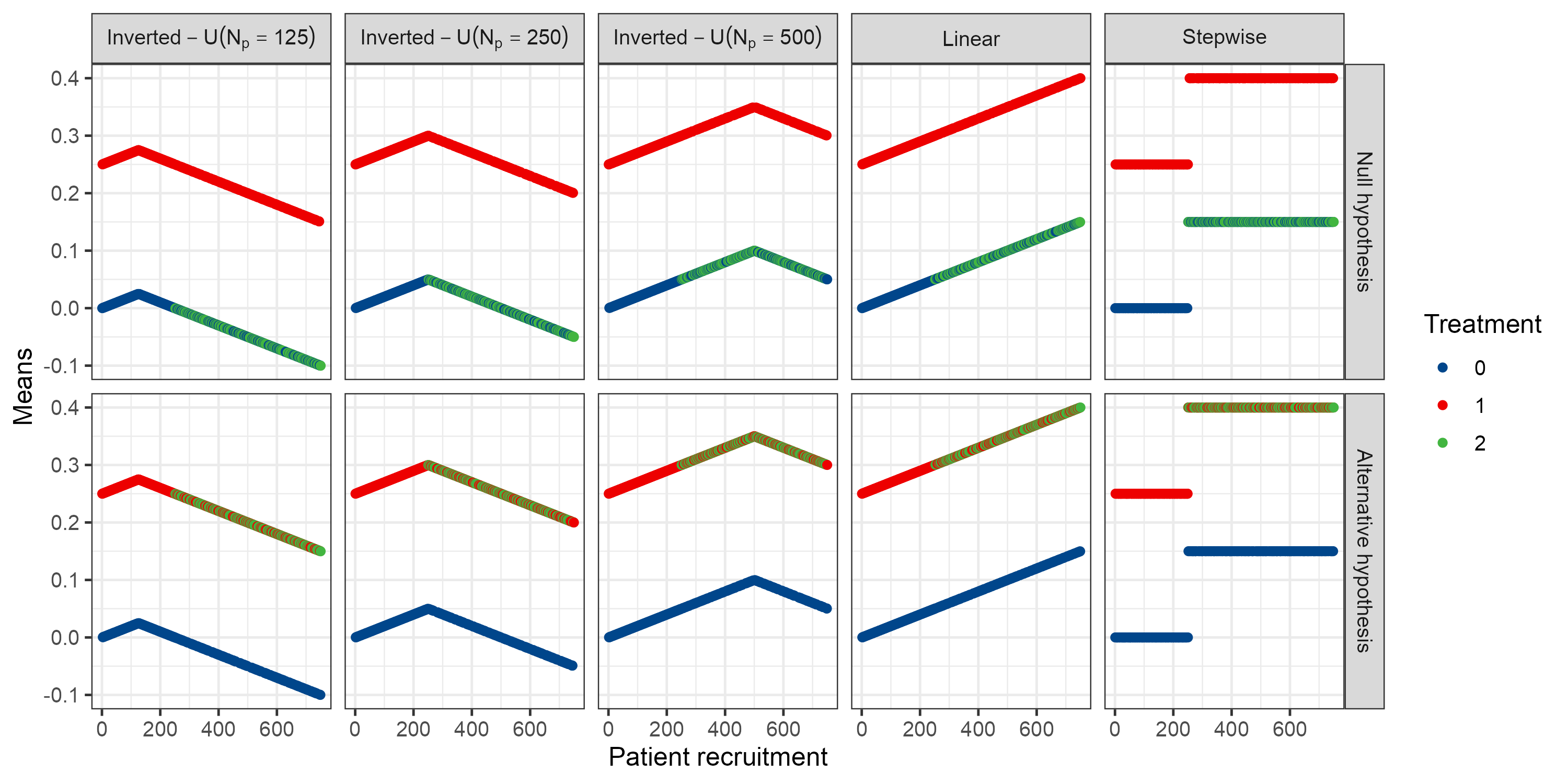}
	\caption{Means for continuous endpoints with respect to 
		patient enrolment to the trial 
		in the presence of equal time trends across arms, depending on the time trends' pattern (linear, step-wise and inverted-U time trends) and for $\lambda_0 = \lambda_1 = \lambda_2 = 0.15$.}
	\label{fig:cont_response_main}
\end{figure}

\clearpage

\newpage



\section*{Supplementary material}

# Supplementary material for "On model-based time trend adjustments in platform trials with non-concurrent controls"

Marta Bofill Roig, Pavla Krotka, Carl-Fredrik Burman, Ekkehard Glimm, Stefan M. Gold, Katharina Hees, Peter Jacko, Franz Koenig, Dominic Magirr, Peter Mesenbrink, Kert Viele, and Martin Posch*

# Contents

---

*martin.posch@meduniwien.ac.at

In this supplementary material, we present the derivation of the weights presented in Section 3.1.1 of the main paper (in Section A), together with an example of weights computation (in Section B), and proofs of the properties of the estimators explained in Section 3.1.2 (in Section C). Moreover, we report further remarks on the impact of time trends on the variance estimation in Section D.

We also present here the results of the simulation study and results from additional simulations (in Sections E and F, respectively). Regarding the simulation results in Section E, we show the bias and root mean squared error of the treatment effect estimators in the presence of time trends. We also present the results for those models in Section 4 of the main paper.

Throughout this document, we use the following abbreviations for the models: ALLTC refers to models using all treatment data and control (see equations 1 and 6 in the main paper), ALLTCI to models using all treatment data and control with the interaction between time and treatment arm (see equations 2 and 7), and TC to models using only data from one treatment arm and the control (see equations 5 and 8); the suffixes Step/Linear indicate whether models adjust time linearly or step-wise.

# A Weights derivation

We may obtain the estimates for the linear model by forming the design matrix $X$, with one row for each patient and with columns for the intercept, arm 1, arm 2, and 2nd time period. If the interaction is included, then $X$ contains an additional column for interaction of arm 1 and the 2nd time period. The matrix elements are 0's and 1's, where 1 indicates that a particular patient matches up to a particular feature. For instance, for the model without interaction, we have

$$X = \begin{matrix} n_{0,1} \\ n_{1,1} \\ n_{0,2} \\ n_{1,2} \\ n_{2,2} \end{matrix} \begin{pmatrix} 1 & 0 & 0 & 0 \\ 1 & 1 & 0 & 0 \\ 1 & 0 & 0 & 1 \\ 1 & 1 & 0 & 1 \\ 1 & 0 & 1 & 1 \end{pmatrix}$$

where the row labels indicate their multiplicities.

The vector estimate of all parameters is $(X^\mathsf{T} X)^{-1} X^\mathsf{T} y$, where $y$ is the vector of patient responses. Thus each parameter estimate is a linear combination of the observations. Our interest lies in the estimate of the arm 2 treatment effect

$$\hat{\theta}_2 = \sum_j v_j y_j$$

where $v_j$ is the $[3, j]$ element of $(X^\mathsf{T} X)^{-1} X^\mathsf{T}$ (since row 3 corresponds to the treatment effect of arm 2) and $y_j$ is the response for the $j^{th}$ patient. Note that all patients with the same covariate values (and thus identical rows of the $X$ matrix) will have the same $v_j$ values, allowing us to collect terms and rewrite

$$\hat{\theta}_2 = \sum_{k,s} w_{k,s} \bar{y}_{k,s}$$

where $w_{k,s} = n_{k,s} v_{k,s}$ and $v_{k,s}$ is the common value of $v_j$ for all patients in arm $k$ and time period $s$. Note that the weighted average is determined entirely by the sample sizes in each arm and time period, not the responses for those patients.

If there are no arm 2 patients in time period 1 ($n_{2,1} = 0$), and the interaction model is used, then the weighted average is extremely simple, with all $w_{k,s} = 0$ except for $w_{0,2} = -1$ and $w_{2,2} = 1$. (We omit the mathematical derivation.) In other words, the estimated treatment is simply the difference in means in the 2nd time period, resulting in the concurrent controls analysis.

If there are no arm 2 patients in time period 1 ($n_{2,1} = 0$), and the non-interaction model is used, then the weights can be obtained as follows. Let $\boldsymbol{v}$ be a column vector created by concatenating individual patient weights $v_j$ so that the first $n_{0,1}$ elements correspond to patients on arm 0 in period 1, the following $n_{1,1}$ elements correspond to patients on arm 1 in period 1, the following $n_{0,2}$ elements correspond to patients on arm 0 in period 2, the following $n_{1,2}$ elements correspond to patients on arm 1 in period 2, the final $n_{2,2}$

elements correspond to patients on arm 2 in period 2. Since the 3rd row of $(X^{\mathsf{T}}X)^{-1}X^{\mathsf{T}}$ corresponds to row vector $\boldsymbol{v}^{\mathsf{T}}$, after taking a transpose we have that the 3rd column of $X(X^{\mathsf{T}}X)^{-1}$ corresponds to vector $\boldsymbol{v}$. Also, observe that

$$
\begin{array}{ccccc}
{\scriptstyle n_{0,1}} & {\scriptstyle n_{1,1}} & {\scriptstyle n_{0,2}} & {\scriptstyle n_{1,2}} & {\scriptstyle n_{2,2}}
\end{array}
\begin{pmatrix}
1 & 0 & 0 & 0 & 0 \\
0 & 1 & 0 & 0 & 0 \\
0 & 0 & 1 & 0 & 0 \\
0 & 0 & 0 & 1 & 0 \\
0 & 0 & 0 & 0 & 1
\end{pmatrix}
\boldsymbol{v} =
\begin{pmatrix}
w_{0,1} \\
w_{1,1} \\
w_{0,2} \\
w_{1,2} \\
w_{2,2}
\end{pmatrix}
$$

where the column labels indicate column multiplicities. Then,

$$
X(X^{\mathsf{T}}X)^{-1}
\begin{pmatrix}
0 \\
0 \\
1 \\
0
\end{pmatrix}
= \boldsymbol{v}
$$

is equivalent to

$$
\begin{array}{ccccc}
{\scriptstyle n_{0,1}} & {\scriptstyle n_{1,1}} & {\scriptstyle n_{0,2}} & {\scriptstyle n_{1,2}} & {\scriptstyle n_{2,2}}
\end{array}
\begin{pmatrix}
1 & 0 & 0 & 0 & 0 \\
0 & 1 & 0 & 0 & 0 \\
0 & 0 & 1 & 0 & 0 \\
0 & 0 & 0 & 1 & 0 \\
0 & 0 & 0 & 0 & 1
\end{pmatrix}
X(X^{\mathsf{T}}X)^{-1}
\begin{pmatrix}
0 \\
0 \\
1 \\
0
\end{pmatrix}
=
\begin{pmatrix}
w_{0,1} \\
w_{1,1} \\
w_{0,2} \\
w_{1,2} \\
w_{2,2}
\end{pmatrix}
$$

It is easy to see that

$$
\begin{array}{ccccc}
{\scriptstyle n_{0,1}} & {\scriptstyle n_{1,1}} & {\scriptstyle n_{0,2}} & {\scriptstyle n_{1,2}} & {\scriptstyle n_{2,2}}
\end{array}
\begin{pmatrix}
1 & 0 & 0 & 0 & 0 \\
0 & 1 & 0 & 0 & 0 \\
0 & 0 & 1 & 0 & 0 \\
0 & 0 & 0 & 1 & 0 \\
0 & 0 & 0 & 0 & 1
\end{pmatrix}
X =
\begin{pmatrix}
n_{0,1} & 0 & 0 & 0 \\
n_{1,1} & n_{1,1} & 0 & 0 \\
n_{0,2} & 0 & 0 & n_{0,2} \\
n_{1,2} & n_{1,2} & 0 & n_{1,2} \\
n_{2,2} & 0 & n_{2,2} & n_{2,2}
\end{pmatrix}
$$

and that

$$
X^{\mathsf{T}}X =
\begin{pmatrix}
N & n_{1,1}+n_{1,2} & n_{2,2} & n_{0,2}+n_{1,2}+n_{2,2} \\
n_{1,1}+n_{1,2} & n_{1,1}+n_{1,2} & 0 & n_{1,2} \\
n_{2,2} & 0 & n_{2,2} & n_{2,2} \\
n_{0,2}+n_{1,2}+n_{2,2} & n_{1,2} & n_{2,2} & n_{0,2}+n_{1,2}+n_{2,2}
\end{pmatrix}
$$

It can be shown that

$$
(X^{\mathsf{T}}X)^{-1}
\begin{pmatrix}
0 \\
0 \\
1 \\
0
\end{pmatrix}
=
\begin{pmatrix}
-\frac{1}{n_{0,1}}\varrho \\
\left(\frac{1}{n_{0,1}}+\frac{1}{n_{1,1}}\right)\varrho \\
\frac{1}{n_{2,2}}+\left(\frac{1}{n_{0,1}}+\frac{1}{n_{1,1}}+\frac{1}{n_{1,2}}\right)\varrho \\
-\left(\frac{1}{n_{1,1}}+\frac{1}{n_{1,2}}\right)\varrho
\end{pmatrix}
$$

where

$$
\varrho = \frac{\frac{1}{n_{0,2}}}{\frac{1}{n_{0,1}}+\frac{1}{n_{1,1}}+\frac{1}{n_{0,2}}+\frac{1}{n_{1,2}}}
$$

Therefore,

$$
\begin{pmatrix}
w_{0,1} \\
w_{1,1} \\
w_{0,2} \\
w_{1,2} \\
w_{2,2}
\end{pmatrix}
=
\begin{pmatrix}
-\varrho \\
\varrho \\
\varrho - 1 \\
-\varrho \\
1
\end{pmatrix}
$$

# B  Example weights for a particular case

If equal randomisation is employed also for arm 2, that is, $n_{0,2} = n_{1,2} = n_{2,2}$, then we can express the weights using the ratio of period sample sizes. Denoting by $\varphi = N_1/N_2$ and taking into account that $\omega = \frac{3}{2}\varphi$, we can rewrite the weights' matrix as

| $k \backslash s$ | 1 | 2 |
|---|---|---|
| 0 | $-\frac{1}{2} + \frac{1}{3\varphi+2}$ | $-\frac{1}{2} - \frac{1}{3\varphi+2}$ |
| 1 | $\frac{1}{2} - \frac{1}{3\varphi+2}$ | $-\frac{1}{2} + \frac{1}{3\varphi+2}$ |
| 2 | 0 | 1 |

thus obtaining $\tilde{\theta}_2 = (\bar{y}_{2,2} - \bar{y}_{0,2}) + \left(\frac{1}{2} - \frac{1}{3\varphi+2}\right)[(\bar{y}_{1,1} - \bar{y}_{0,1}) - (\bar{y}_{1,2} - \bar{y}_{0,2})]$. For instance, if $N_1/N_2 = 2/3$, then we have

| $k \backslash s$ | 1 | 2 |
|---|---|---|
| 0 | $-0.25$ | $-0.75$ |
| 1 | $0.25$ | $-0.25$ |
| 2 | 0 | 1 |

, and $\tilde{\theta}_2 = (\bar{y}_{2,2} - \bar{y}_{0,2}) + 0.25\,[(\bar{y}_{1,1} - \bar{y}_{0,1}) - (\bar{y}_{1,2} - \bar{y}_{0,2})]$;

# C  Properties of the estimators and tests under model-based period-wise adjustments

Model (4) in the paper, can be written as

$$E(g(Y_j)\,|t_j) = \theta_{k_j} + f(t_j) \tag{1}$$

where $k_j = 0, 1, 2$ is the treatment that patient $j$ was randomised to. We set $\theta_0 := \eta_0$ for convenience of notation. Here, and in all subsequent derivations, we always condition on the observed $k_j$, but treat $t_j$ as random.

We are considering 2 periods: $P = 1$ with $k_j = 0, 1$ and $P = 2$ with $k_j = 0, 1, 2$. Within period, $t_j$ does not depend on $k_j$ due to randomisation. For period 2 alone, we use the model

$$E(g(Y_j)\,|P = 2) = \mu_0 + \nu + \theta_{k_j}. \tag{2}$$

This arises from the conditional model in (1) via

$$E(y_j\,|P = 2) = E(E(y_j\,|t_j, P = 2)) = E(f(t_j)|P = 2) + \theta_{k_j} = \mu_0 + \nu + \theta_{k_j}.$$

Here, $E(f(t_j)|P = 2)$ is the same for all $j$ due to randomisation and $E(f(t_j)|P = 2) =: \mu_0 + \nu$ is just a naming convention.

For period 1 alone, we use the model

$$E(g(Y_j)\,|P = 1) = \mu_0 + \theta_{k_j}. \tag{3}$$

This arises from the conditional model in (1) via

$$E(g(Y_j)\,|P = 1) = E(E(g(Y_j)\,|t_j, P = 1)) = E(f(t_j)|P = 1) + \theta_{k_j} = \mu_0 + \theta_{k_j}.$$

Here, $E(f(t_j)|P = 1)$ is the same for all $j$ due to randomisation and $E(f(t_j)|P = 1) =: \mu_0$ is a naming convention. Taken together, we have $\mu_0 = E(f(t_j)|P = 1)$ and $\nu = E(f(t_j)|P = 2) - E(f(t_j)|P = 1)$.

From these considerations, it is obvious that (if we are not interested in the functional form of $f(t_j)$), we can use the "unconditional-within-period" model characterised by (3) and (2) for estimation of $\theta_{k_j}$ in the conditional model (1). The estimate is unbiased in the linear case where $g()$ is the identity function. In the logistic regression model, this estimate is asymptotically unbiased - just like the point estimates from a correctly specified model. Regarding precision, however, it may be advantageous to estimate $f(t_j)$ to refine the estimate of $\theta_{k_j}$.

Regarding variance estimation, the asymptotic unbiasedness of the parameter estimates in the logistic regression model carries over to the variance estimates, since these are also functions of the estimated response probabilities. In the linear model, where the variance is modelled by an additional separate parameter, however, biases in variance estimation can arise if the residual variance changes in time and the randomisation ratio between control and test treatment changes in time. Additional remarks on these topics can be found in the next section.

# D Impact of time trends on the variance estimation in the linear model

We assume that within period ($P = q$), the responses follow the model $y_j | T = t_j, P = q \sim N(f(t_j) + \theta_{k_j}, \sigma^2_{y \cdot T}(t_j))$. Hence, the variance is considered a function of time. We also assume that within period $q$, the time $t_j$ at which patient $j$'s response $y_j$ is observed, is an independent, identically distributed (i.i.d.) sample from $T_q$. This is fulfilled if patients randomly enter the study at any point in time during the recruitment period and are randomized to a treatment. The distribution of $T_q$ does not depend on the treatment $k_j$, but otherwise we leave it as unspecified. In particular, it might be different for $P = 1$ and $P = 2$. We have

$$E(y_j | P = q) = E(E(y_j | T_q, P = q)) = E(f(T_q)) + \theta_{k_j} =: \mu_q + \theta_{k_j}$$

and

$$var(y_j | P = q) = var(E(y_j | T_q, P = q)) + E(var(y_j | T_q, P = q)) =$$
$$var(f(T_q) + \theta_{k_j}) + E(\sigma^2_{y \cdot T}(T_q)) = var(f(T_q)) + E\left(\sigma^2_{y \cdot T}(T_q)\right).$$

Thus, with $\sigma^2_q = var(f(T_q)) + E\left(\sigma^2_{y \cdot T}(T_q)\right)$, we have that

$$y_j | P = q \sim N(\mu_q + \theta_{k_j}, \sigma^2_q).$$

Within period $q$, the condition that $y_j | P = q$ are i.i.d. normally distributed is met. For $P = 1$ and $P = 2$, however, we have different variances. This can be addressed by fitting the model $y_j | P = q \sim N(\mu_q + \theta_{k_j}, \sigma^2_q)$ with $\mu_q = \mu_0 + \nu \cdot I(P = 2)$.

The (ML- or REML-)estimate $\hat{\theta}_k$ from this model will fulfill that $\sqrt{n}(\hat{\theta}_k - \theta)$ is asymptotically normal. Hence, we see that the step model which accounts for the time trend by introducing a period effect can also account for time-dependent changes in the variance by likewise modeling variance with a step function for period.

Regarding precision of the estimate, there will be a loss relative to knowledge of $f(\cdot)$: If we know $f(\cdot)$, then by conditioning on the observed $t_j$, we can eliminate $var(f(T_q))$ from $\sigma^2_q$.

A curious by-product of the above consideration is that we do not need to know $\sigma^2_{y \cdot T}(t_j)$ either. In practice, it might often be assumed that this does not depend on $t_j$. If we know that $\sigma^2_{y \cdot T}(t_j) = \sigma^2_{y \cdot T}$ and also know $f(\cdot)$, then $\sigma^2 = \sigma^2_q = \sigma^2_{y \cdot T}$ arises. If we do not know $f(\cdot)$, however, the assumption $\sigma^2_{y \cdot T}(t_j) = \sigma^2_{y \cdot T}$ does not prevent $\sigma^2_q$ from being different for $q = 1$ and $q = 2$, so we might as well relax this assumption.

Let us now consider the case where it is (possibly erroneously) assumed that the residual variance is the same in the two periods. Hence, we have stochastically independent observations $y_j \sim N\left(\mathbf{x}'_j \boldsymbol{\theta}, \sigma^2_j\right)$, $j = 1, \ldots, n$. For the sake of estimation, it is assumed that $\sigma^2_j = \sigma^2$, but in reality, they might be different, e.g. depend on period of recruitment.

Let $\mathbf{X} = (\mathbf{x}_1, \ldots, \mathbf{x}_n)'$ be the $n \times p$-design matrix. We estimate $\boldsymbol{\theta}$ by $\hat{\boldsymbol{\theta}} = (\mathbf{X}'\mathbf{X})^{-1} \mathbf{X}'\mathbf{y}$ such that

$$\hat{\boldsymbol{\theta}} \sim N\left(\boldsymbol{\theta}, (\mathbf{X}'\mathbf{X})^{-1} \mathbf{X}'\boldsymbol{\Sigma}\mathbf{X} (\mathbf{X}'\mathbf{X})^{-1}\right)$$

where $\boldsymbol{\Sigma} = Diag(\sigma^2_j)_{j=1,\ldots,n}$. The residual error variance is estimated by

$$s^2 = \frac{1}{n}\mathbf{y}' \left(\mathbf{I}_n - \mathbf{X} (\mathbf{X}'\mathbf{X})^{-1} \mathbf{X}'\right) \mathbf{y} = \frac{1}{n}\sum_{j=1}^{n} \left(y_j - \mathbf{x}'_j \hat{\boldsymbol{\theta}}\right)^2.$$

If $n \to \infty$, the residuals $y_j - \mathbf{x}_j'\hat{\boldsymbol{\theta}}$ become asymptotically stochastically independent with variance $\sigma_j^2$ and hence $s^2$ converges to $\frac{1}{n}\sum_{j=1}^n \sigma_j^2 =: \bar{\sigma}^2$.

Assume now that we want to test a hypothesis $H_0 : \mathbf{h}'\boldsymbol{\theta} = 0$. We use

$$t = \frac{\mathbf{h}'\hat{\boldsymbol{\theta}}}{\sqrt{s^2 \cdot \mathbf{h}' \left(\mathbf{X}'\mathbf{X}\right)^{-1} \mathbf{h}}}.$$

Since the true variance of $\mathbf{h}'\hat{\boldsymbol{\theta}}$ is $\mathbf{h}' \left(\mathbf{X}'\mathbf{X}\right)^{-1} \mathbf{X}'\boldsymbol{\Sigma}\mathbf{X} \left(\mathbf{X}'\mathbf{X}\right)^{-1} \mathbf{h}$, $t \sim N(0,1)$ holds asymptotically under $H_0$ if

$$\bar{\sigma}^2 \cdot \mathbf{h}' \left(\mathbf{X}'\mathbf{X}\right)^{-1} \mathbf{h} = \mathbf{h}' \left(\mathbf{X}'\mathbf{X}\right)^{-1} \mathbf{X}'\boldsymbol{\Sigma}\mathbf{X} \left(\mathbf{X}'\mathbf{X}\right)^{-1} \mathbf{h}. \tag{4}$$

Hence, if we are able to find a parametrization of the model in which we show that

$$\mathbf{X}'\boldsymbol{\Sigma}\mathbf{X} = \bar{\sigma}^2 \cdot \mathbf{X}'\mathbf{X}, \tag{5}$$

then $t \sim N(0,1)$ holds asymptotically under $H_0$.

Since $\boldsymbol{\Sigma}$ is a diagonal matrix, we have $\mathbf{X}'\boldsymbol{\Sigma}\mathbf{X} = \sum_{j=1}^n \sigma_j^2 \cdot \mathbf{x}_j \mathbf{x}_j'$ and $\mathbf{X}'\mathbf{X} = \sum_{j=1}^n \mathbf{x}_j \mathbf{x}_j'$. A simple example where this holds is the following: Assume a two-group comparison $H_0 : \mu_1 - \mu_0 = 0$. Then

$$\mathbf{X} = \left( \begin{array}{cc} \mathbf{1}_{n_0} & \mathbf{0}_{n_0} \\ \mathbf{0}_{n_1} & \mathbf{1}_{n_1} \end{array} \right).$$

Furthermore, $\mathbf{x}_j \mathbf{x}_j' = \left( \begin{array}{cc} 1 & 0 \\ 0 & 0 \end{array} \right)$ for a patient in group 0 and $\mathbf{x}_j \mathbf{x}_j' = \left( \begin{array}{cc} 0 & 0 \\ 0 & 1 \end{array} \right)$ for a patient in group 1. Thus, $\mathbf{X}'\mathbf{X} = \left( \begin{array}{cc} n_0 & 0 \\ 0 & n_1 \end{array} \right)$ and

$$\mathbf{X}'\boldsymbol{\Sigma}\mathbf{X} = \left( \sigma_1^2 + \ldots + \sigma_{n_0}^2 \right) \left( \begin{array}{cc} 1 & 0 \\ 0 & 0 \end{array} \right) + \left( \sigma_{n_0+1}^2 + \ldots + \sigma_n^2 \right) \left( \begin{array}{cc} 0 & 0 \\ 0 & 1 \end{array} \right).$$

Hence, $\mathbf{X}'\boldsymbol{\Sigma}\mathbf{X} = \bar{\sigma}^2 \mathbf{X}'\mathbf{X}$ if

$$(\sigma_1^2 + \ldots + \sigma_{n_0}^2)/n_0 = \left( \sigma_{n_0+1}^2 + \ldots + \sigma_n^2 \right)/n_1 = \bar{\sigma}^2.$$

This example corresponds to the intuition that if we want to compare the means of two groups and these groups share the same average variance, then the $t$-test defined by $t$ asymptotically keeps the type I error. From this consideration, it is also clear that examples where equation (5) does not hold can be constructed, e.g. if the majority of observations on treatment 0 is from period 1, the majority of observations from treatment 1 is from period 2 and the variance is larger in period 2 than in period 1.

# E   Results of the simulation study

All simulations were carried out using 100,000 replications per scenario. At each iteration, all models were fitted to the same simulated datasets. This explains common patterns regarding the smoothness of the estimates with respect to time trends (see, for instance, Figure S6).

## E.1   Continuous endpoints

In what follows, we present additional results for continuous endpoints in the presence of time trends, which are either equal across all arms (Section E.1.1), or equal in the control group and treatment 2 but different in treatment 1 (Section E.1.2). We consider the time trend's patterns: linear, step-wise and inverted-U. For the inverted-U trend, we consider three different settings depending on the point at which the trend changes from positive to negative. This point can be either in the middle of period 1 ($N_p = 125$), between the two periods ($N_p = 250$) or in the middle of period 2 ($N_p = 500$).

## E.1.1   Equal time trends

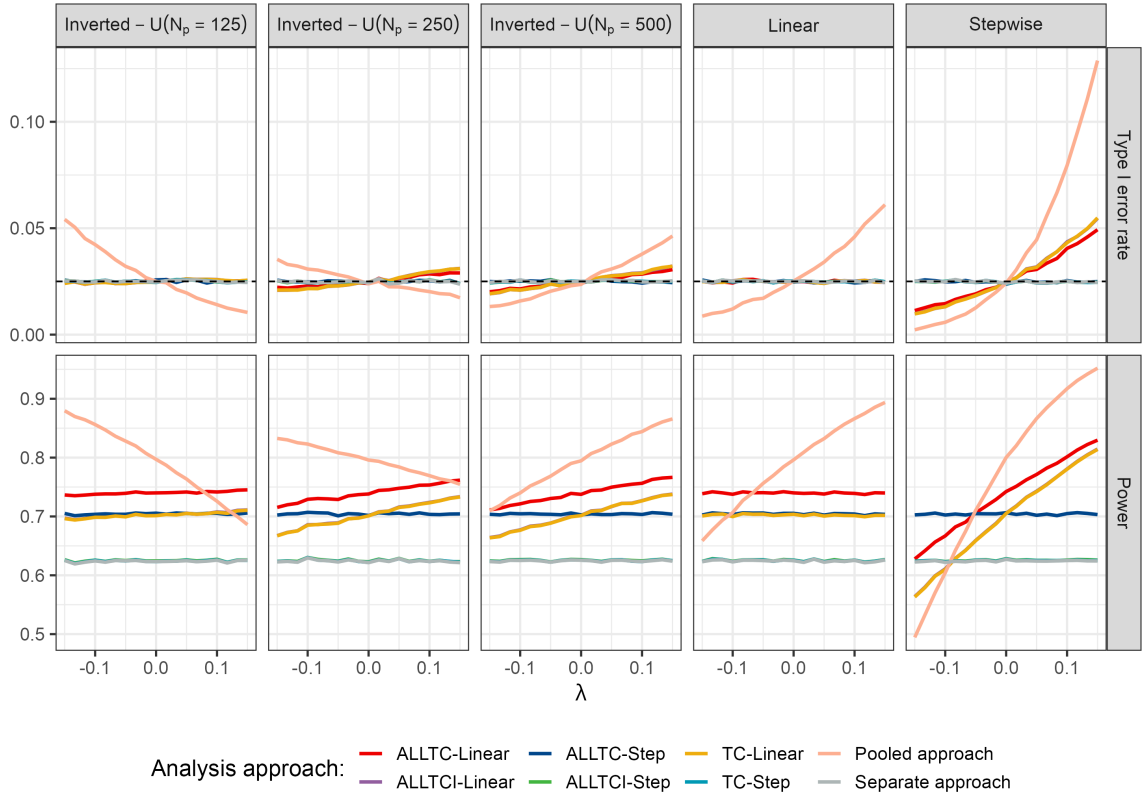

Figure S1: Type I error rate and power of rejecting $H_{02}$ for continuous endpoints in the presence of equal linear, step-wise and inverted-U time trends across arms with respect to the strength of the time trend ($\lambda = \lambda_k$, $k = 0, 1, 2$) and according to the model used. Note that some lines overlap, e.g., ALLTC-Step and the separate approach are overlapping in the figures of the first row; ALLTCI-Step, TC-Step and the separate approach are overlapping in the figures of the second row.

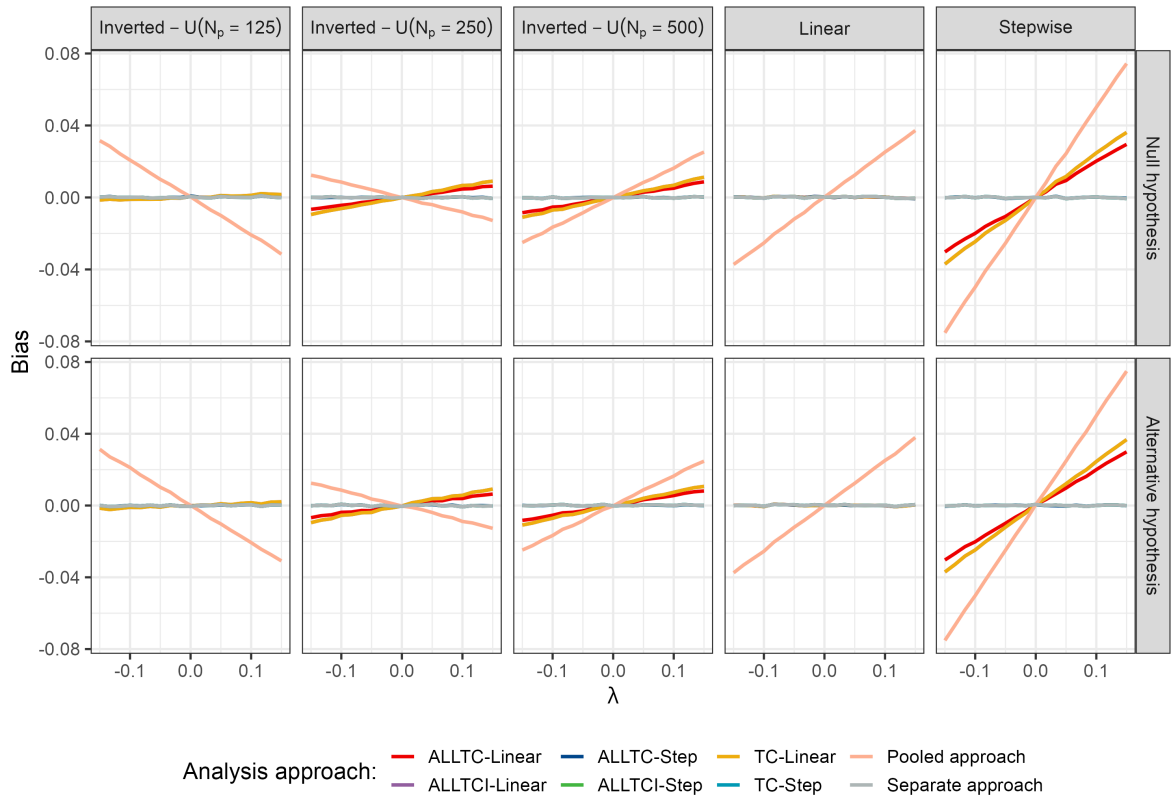

Figure S2: Bias of the treatment effect estimators (difference in means between control arm and treatment 2) for continuous endpoints in the presence of equal linear, step-wise and inverted-U time trends across arms with respect to the strength of the time trend ($\lambda = \lambda_k$, $k = 0, 1, 2$) and according to the model used.

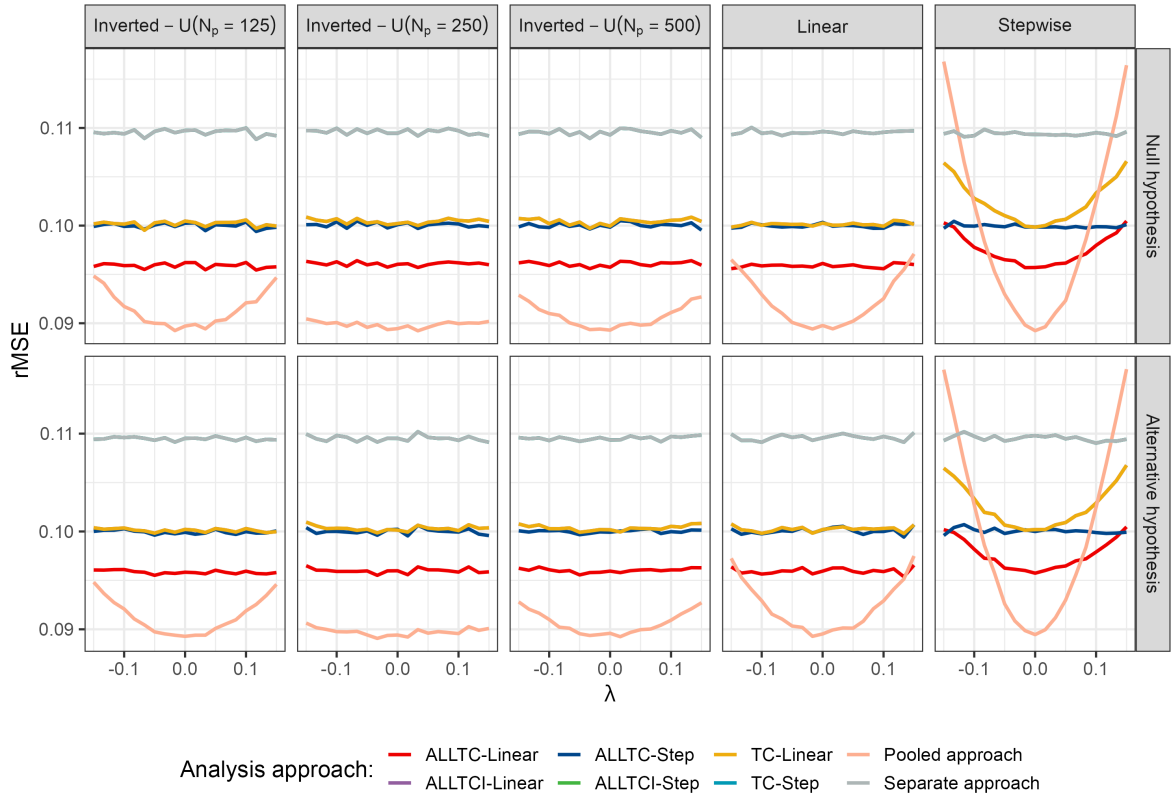

Figure S3: Root mean squared error of the treatment effect estimators (difference in means between control arm and treatment 2) for continuous endpoints in the presence of equal linear, step-wise and inverted-U time trends across arms with respect to the strength of the time trend ($\lambda = \lambda_k$, $k = 0, 1, 2$) and according to the model used.

### E.1.2 Different time trends

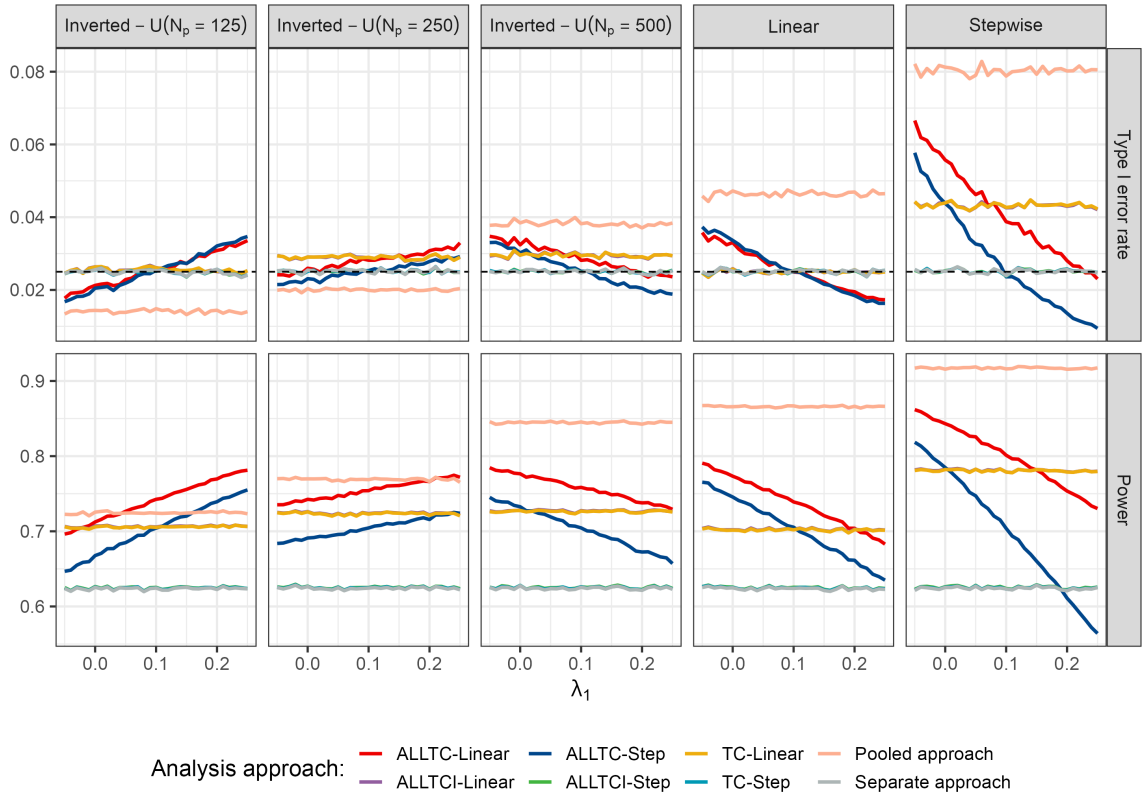

Figure S4: Type I error rate and power of rejecting $H_{02}$ for continuous endpoints in the presence of different linear, step-wise and inverted-U time trends (for $\lambda_0 = \lambda_2 = 0.1$) with respect to the strength of the time trend in treatment arm 1 ($\lambda_1$) and according to the model used. Note that some lines overlap, e.g., ALLTCI-Step, TC-Step and the separate approach are overlapping in the figures of the second row.

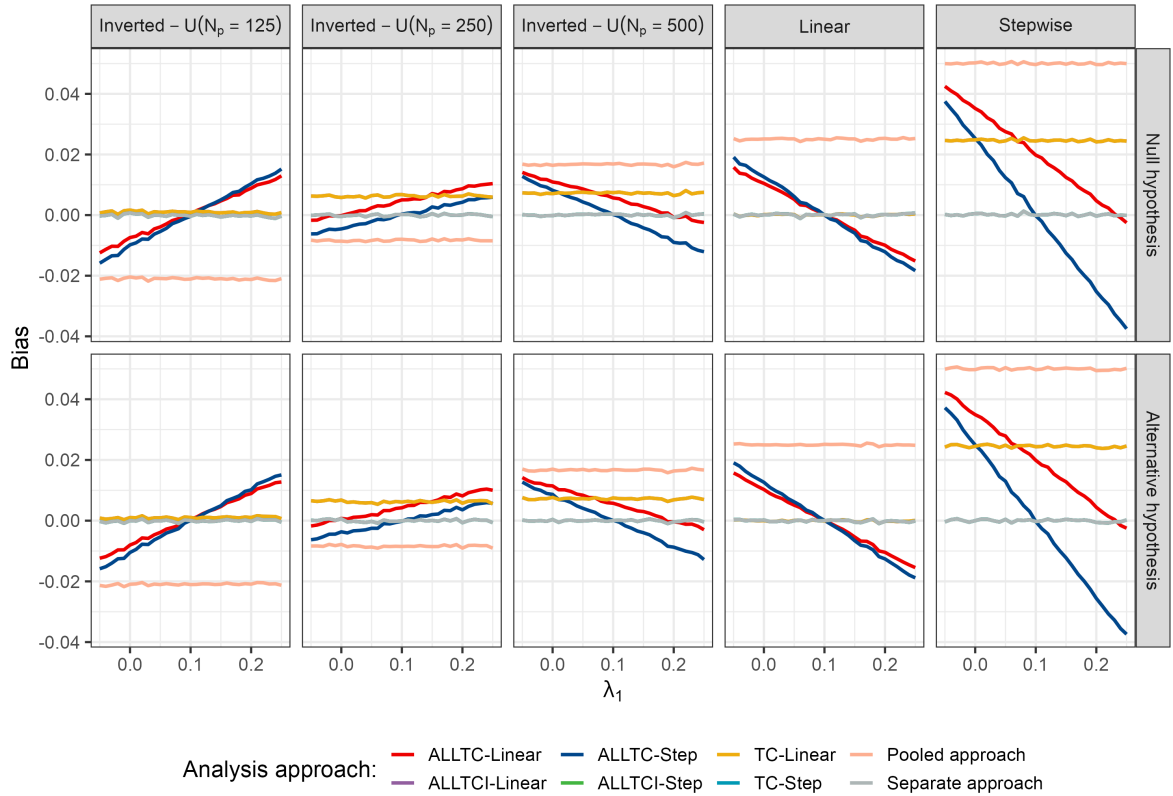

Figure S5: Bias of the treatment effect estimators (difference in means between control arm and treatment 2) for continuous endpoints in the presence of different linear, step-wise and inverted-U time trends (for $\lambda_0 = \lambda_2 = 0.1$) with respect to the strength of the time trend in treatment arm 1 ($\lambda_1$) and according to the model used.

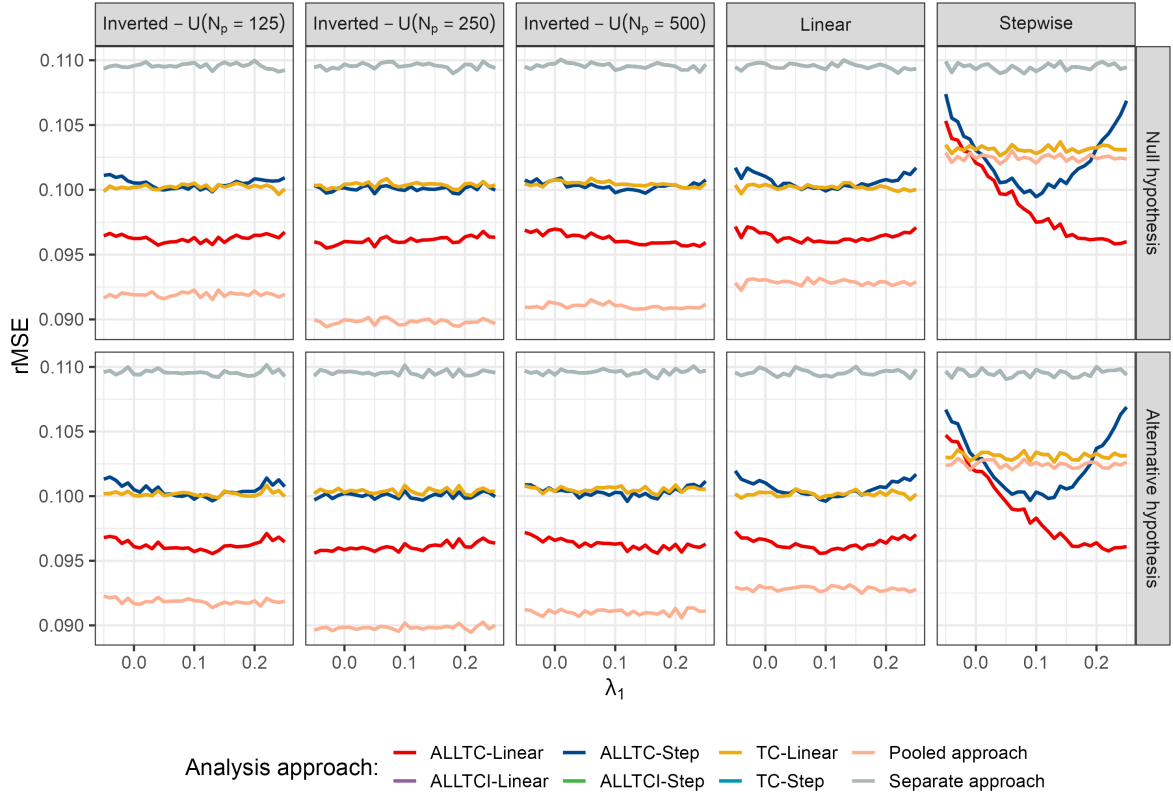

Figure S6: Root mean squared error of the treatment effect estimators (difference in means between control arm and treatment 2) for continuous endpoints in the presence of different linear, step-wise and inverted-U time trends (for $\lambda_0 = \lambda_2 = 0.1$) with respect to the strength of the time trend in treatment arm 1 ($\lambda_1$) and according to the model used.

## E.2  Binary endpoints

Next, we present additional results for binary endpoints in the presence of time trends. As before, time trends can be either equal across all arms (Section E.2.1), or equal in the control group and treatment 2 but different in treatment 1 (Section E.2.2). We consider linear, step-wise and inverted-U time trend's patterns. As before, for the inverted-U trend, the trend switches in the middle of period 1 ($N_p = 125$), between the two periods ($N_p = 250$) or in the middle of period 2 ($N_p = 500$).

For scenarios with different time trends, we considered both, positive and negative trends for the control arm and treatment 2. Here, $\lambda_0 = \lambda_2$ were chosen in order to achieve 5% drop ($\lambda_0 = \lambda_2 < 0$) or increase ($\lambda_0 = \lambda_2 > 0$) in the response rate in the control arm from period 1 to period 2. Results for scenarios with positive time trend in the control and treatment 2 are shown in Figures S10, S12, S14 and S15. For scenarios with negative time trends in the control and treatment 2, the results are presented in Figures S11 and S13.

### E.2.1 Equal time trends

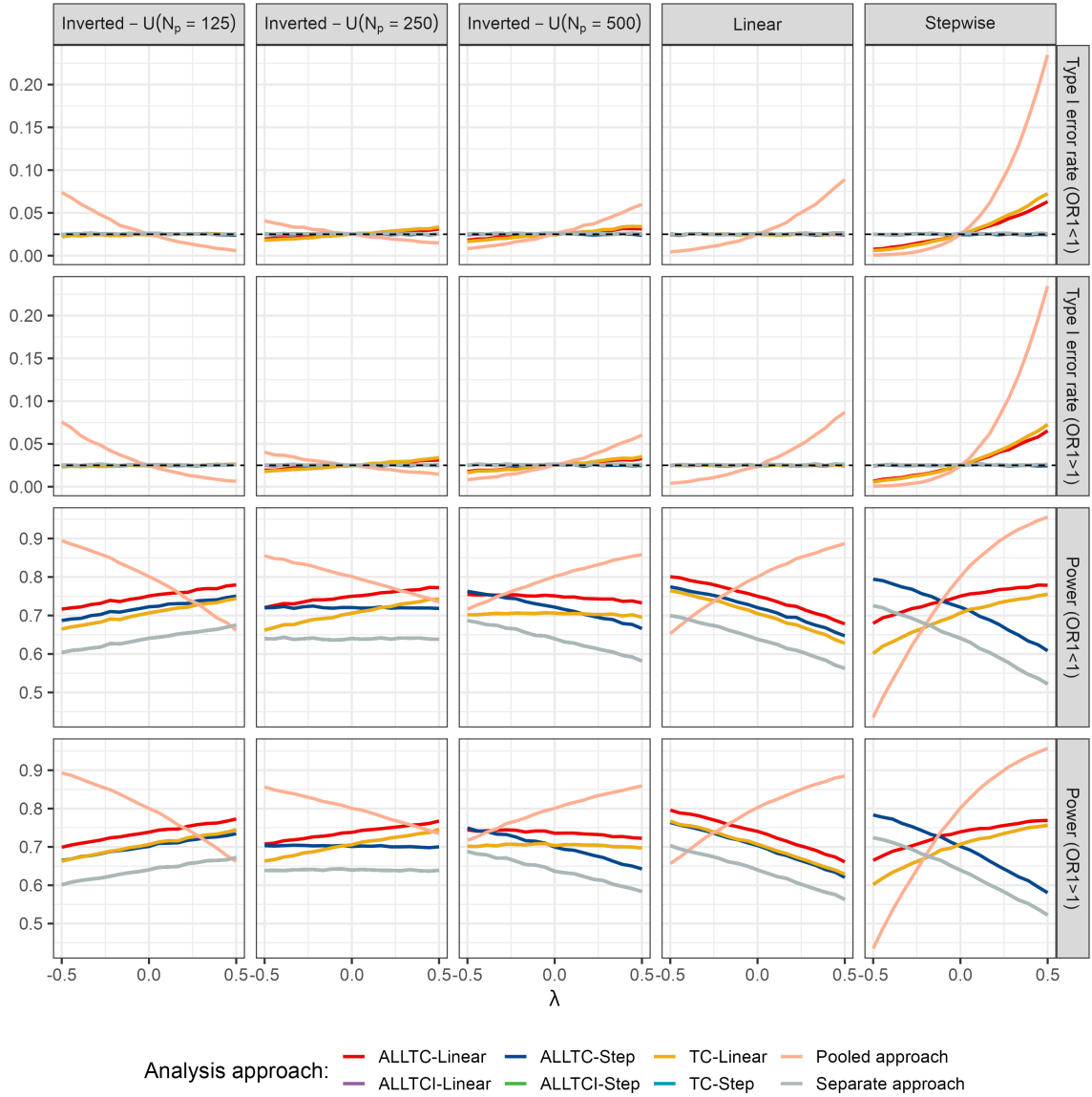

Figure S7: Type I error rate and power of rejecting $H_{02}$ for binary endpoints in the presence of equal linear, step-wise and inverted-U time trends across arms with respect to the strength of the time trend ($\lambda = \lambda_k$, $k = 0, 1, 2$) and according to the model used. Note that some lines overlap, e.g., ALLTC-Step and the separate approach are overlapping in the figures of the first two rows.

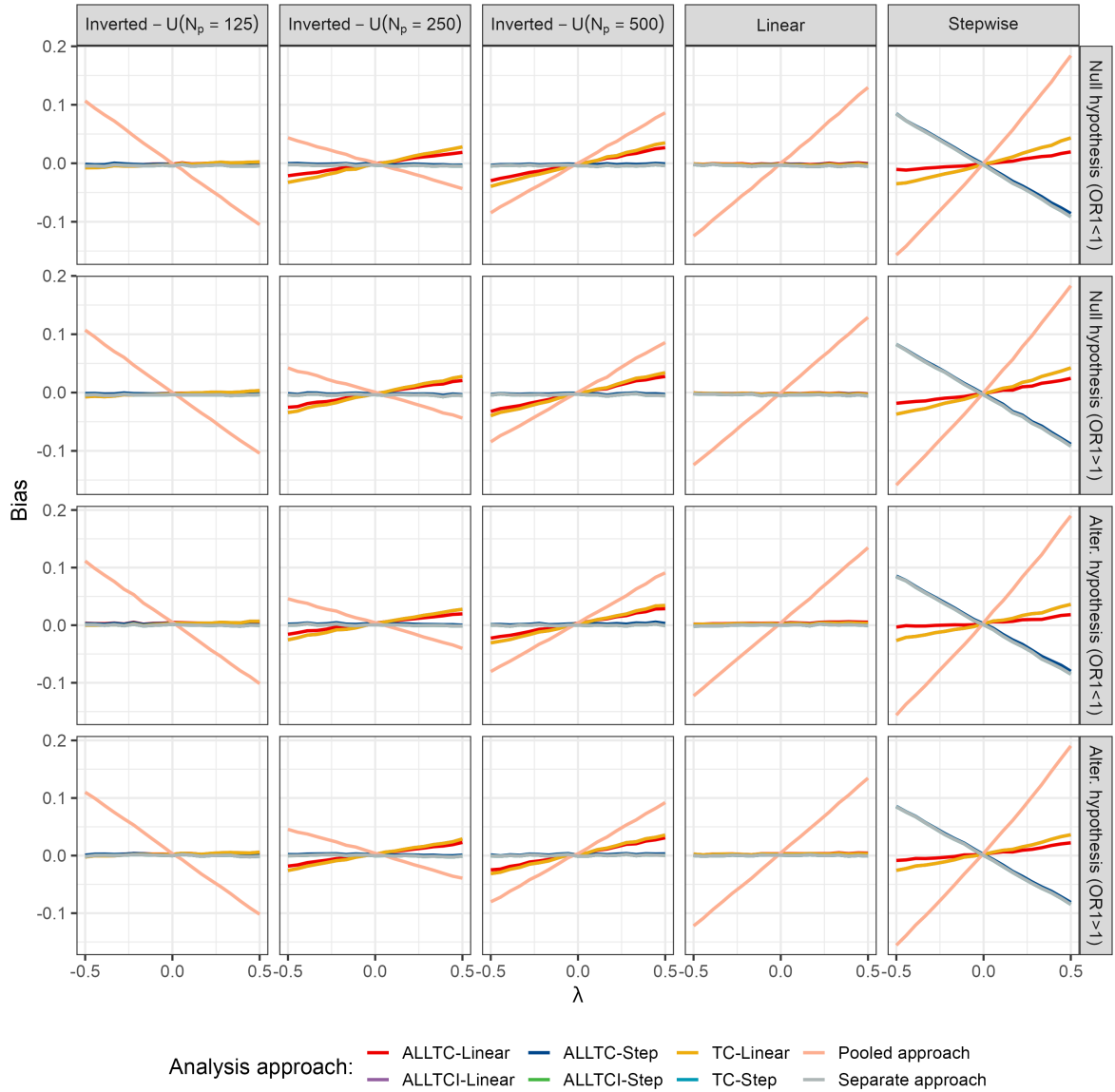

Figure S8: Bias of the treatment effect estimators $(\log(OR_2))$ for binary endpoints in the presence of equal linear, step-wise and inverted-U time trends across arms with respect to the strength of the time trend $(\lambda = \lambda_k, \ k = 0, 1, 2)$ and according to the model used.

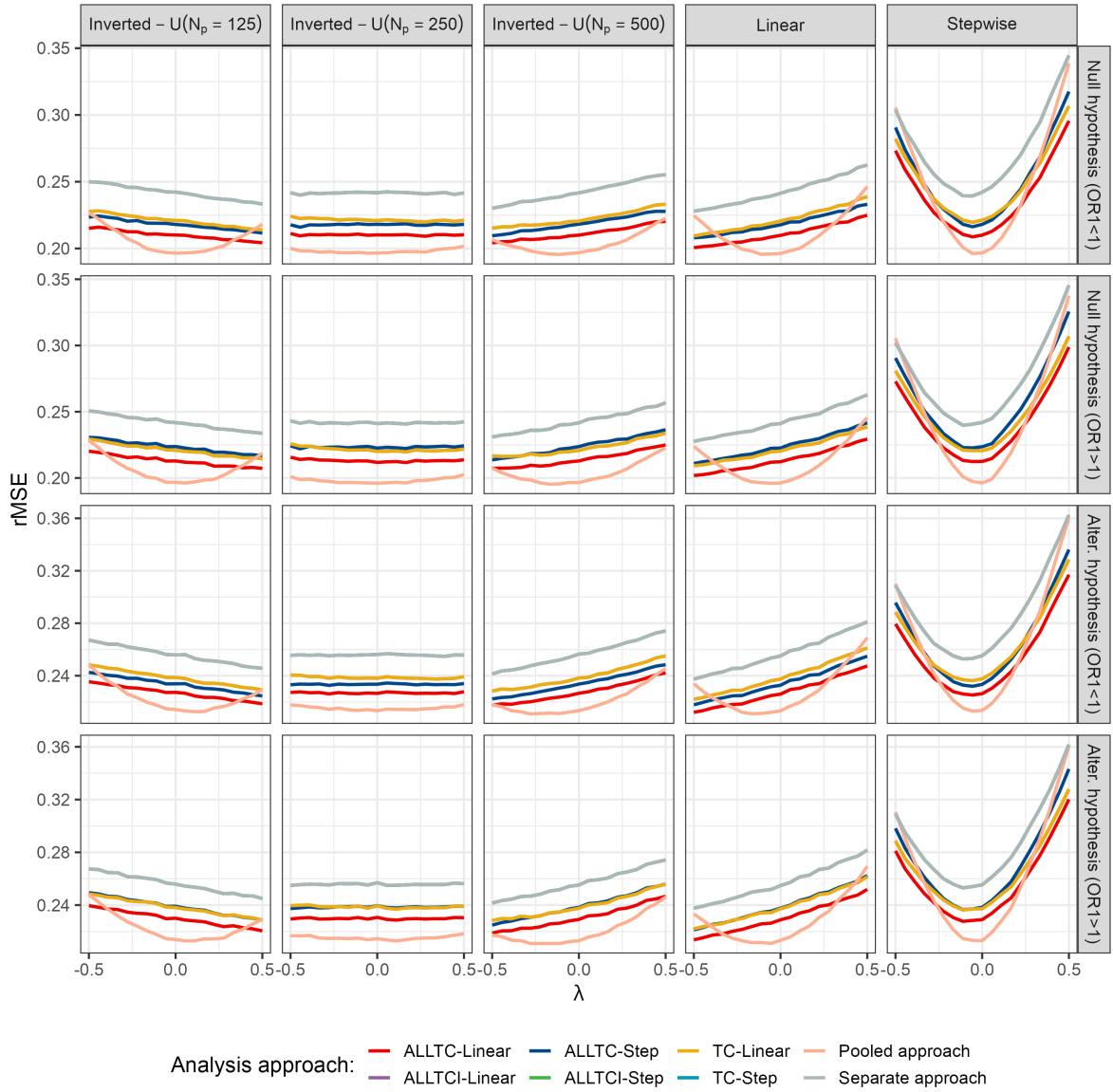

Figure S9: Root mean squared error of the treatment effect estimators $(\log(OR_2))$ for binary endpoints in the presence of equal linear, step-wise and inverted-U time trends across arms with respect to the strength of the time trend ($\lambda = \lambda_k$, $k = 0, 1, 2$) and according to the model used.

## E.2.2 Different time trends

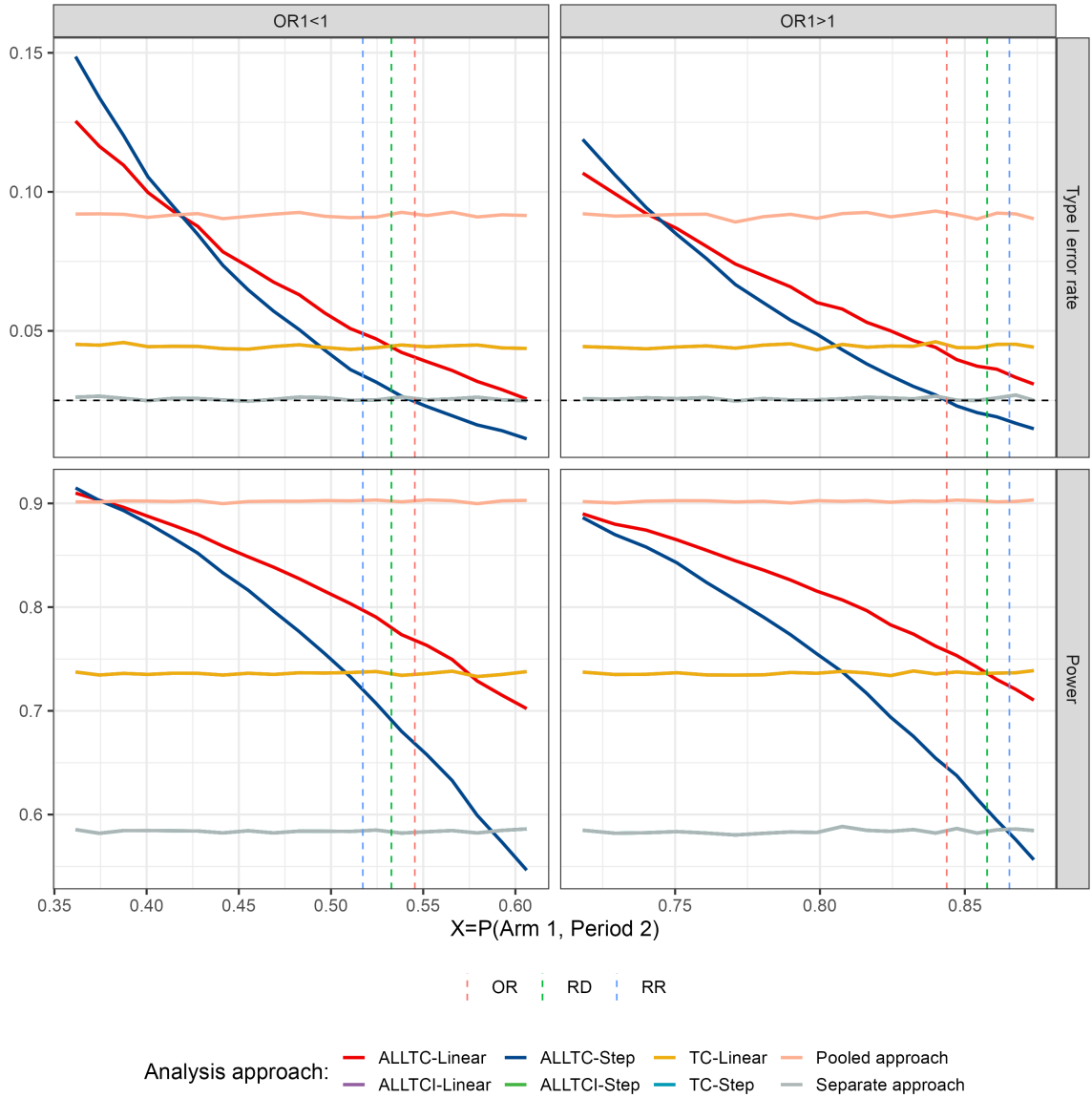

Figure S10: Type I error rate and power of rejecting $H_{02}$ for binary endpoints in the presence of different step-wise time trends when there is a positive time trend in the control ($\lambda_0 = \lambda_2 > 0$ and varying $\lambda_1$) with respect to the response rate in treatment arm 1 in the second period, and depending on the model used. Note that some lines overlap, e.g., ALLTCI-Step, TC-Step and the separate approach are overlapping in the figures of the first and second rows.

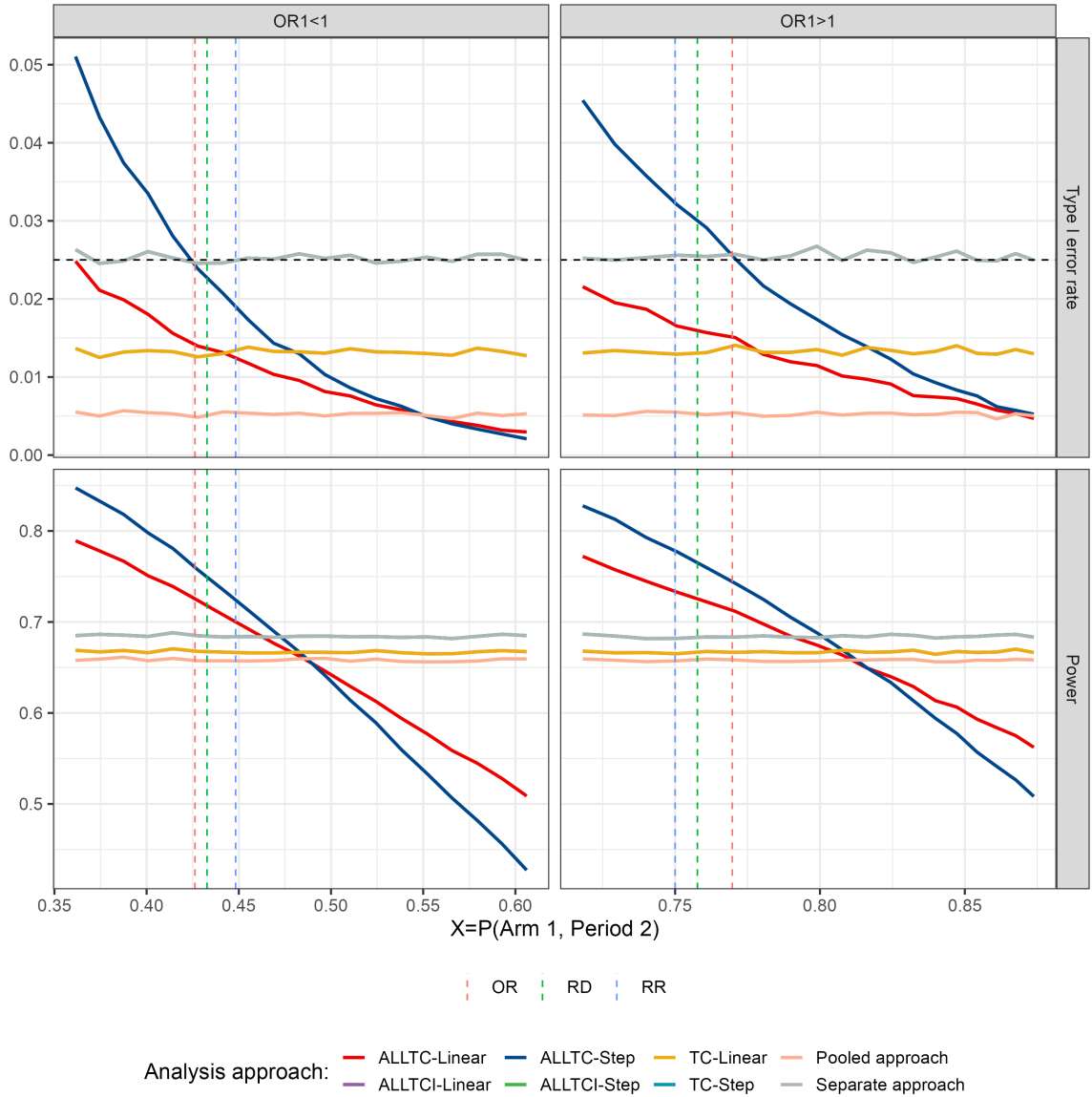

Figure S11: Type I error rate and power of rejecting $H_{02}$ for binary endpoints in the presence of step-wise time trends when there is a negative time trend in the control ($\lambda_0 = \lambda_2 < 0$ and varying $\lambda_1$) with respect to the response rate in treatment arm 1 in the second period, and depending on the model used. Note that some lines overlap, e.g., ALLTCI-Step, TC-Step and the separate approach are overlapping in the figures of the first and second rows.

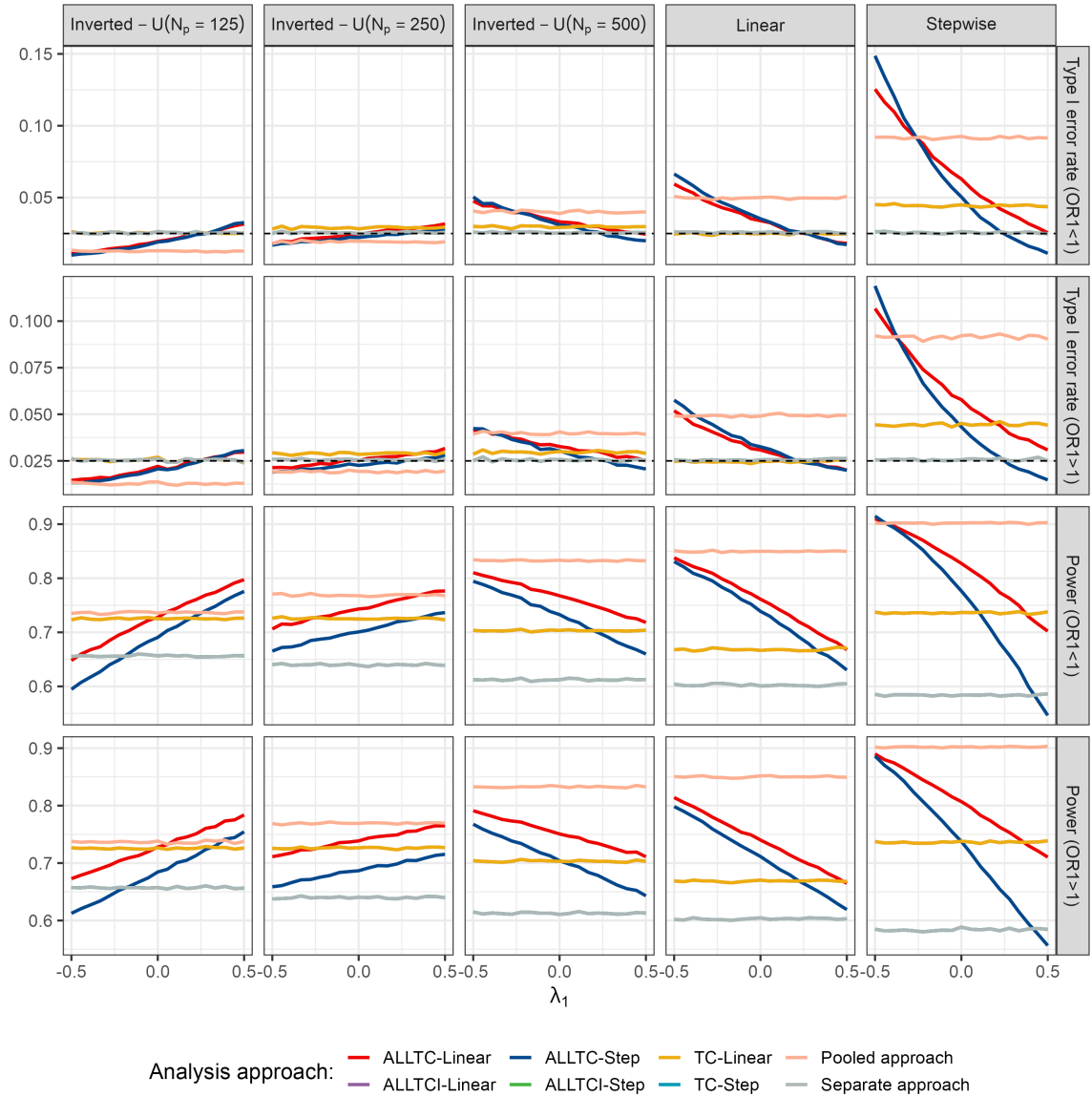

Figure S12: Type I error rate and power of rejecting $H_{02}$ for binary endpoints in the presence of different linear, step-wise and inverted-U time trends when there is a positive time trend in the control ($\lambda_0 = \lambda_2 > 0$) with respect to the strength of the time trend in treatment arm 1 ($\lambda_1$) and according to the model used. Note that some lines overlap, e.g., ALLTCI-Step, TC-Step and the separate approach are overlapping in the figures of the first and second rows.

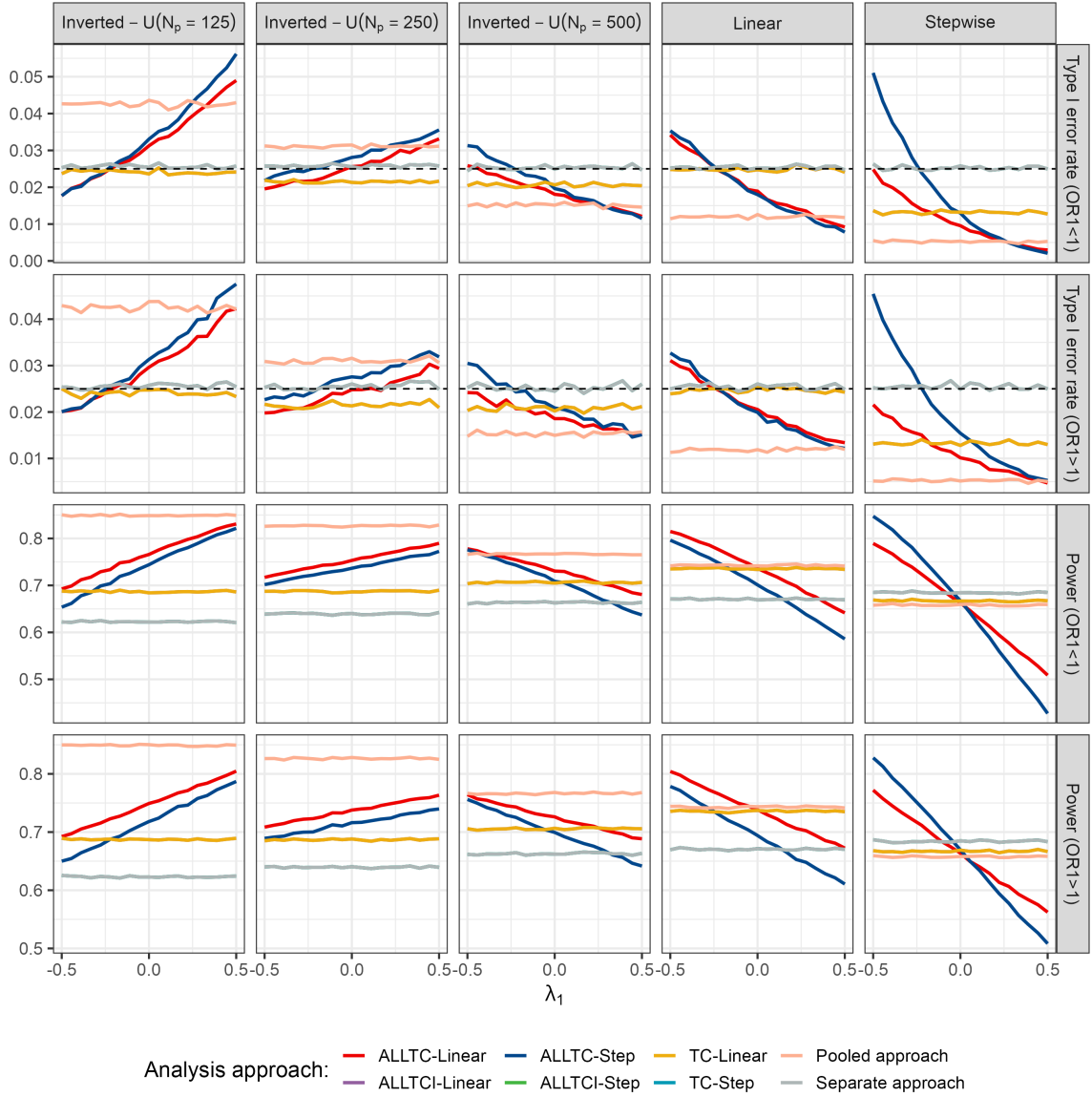

Figure S13: Type I error rate and power of rejecting $H_{02}$ for binary endpoints in the presence of linear, step-wise and inverted-U time trends when there is a negative time trend in the control ($\lambda_0 = \lambda_2 < 0$) with respect to the strength of the time trend in treatment arm 1 ($\lambda_1$) and according to the model used. Note that some lines overlap, e.g., ALLTCI-Step, TC-Step and the separate approach are overlapping in the figures of the first and second rows.

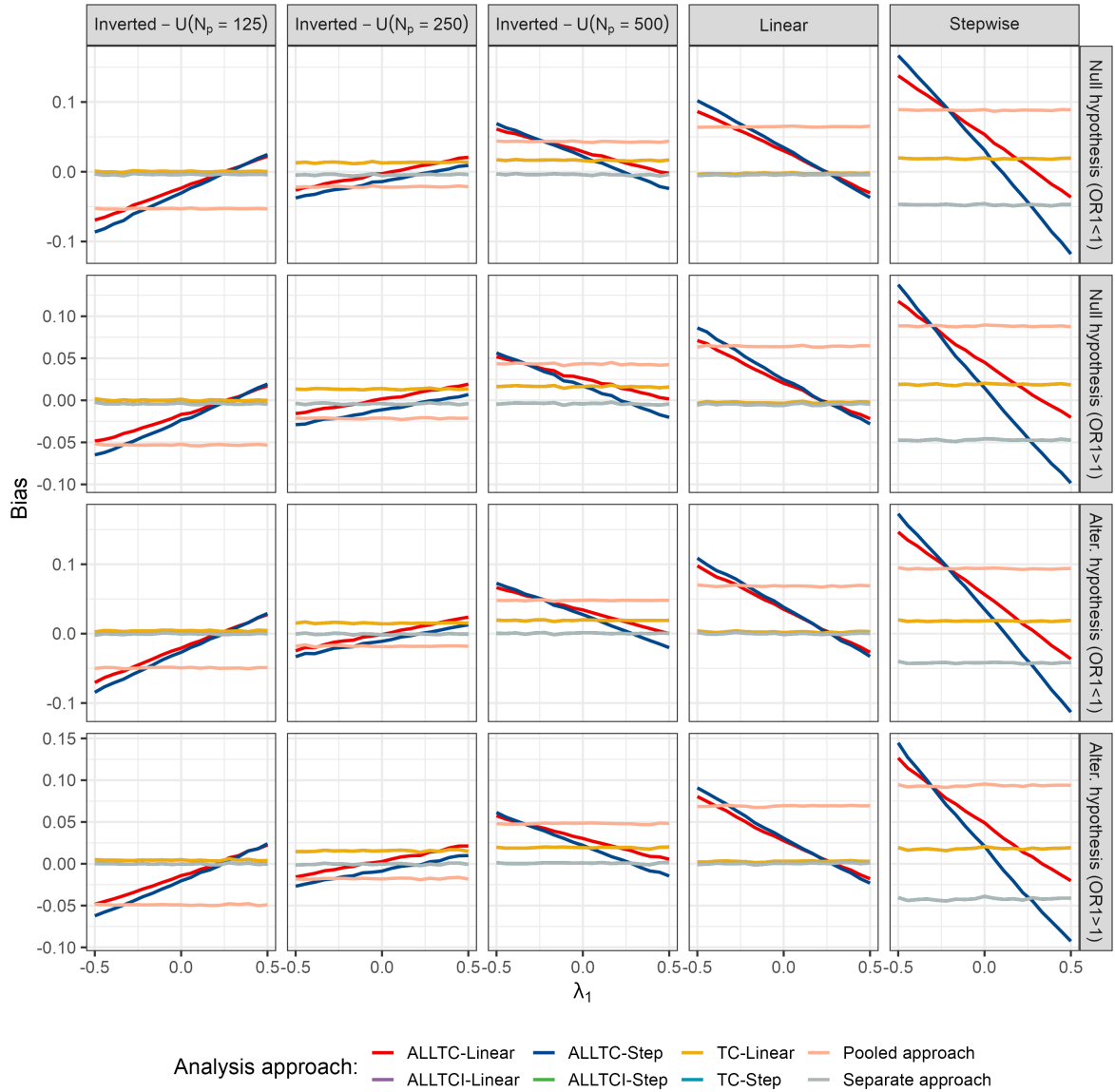

Figure S14: Bias of the treatment effect estimators ($\log(OR_2)$) for binary endpoints in the presence of linear, step-wise and inverted-U time trends when there is a positive time trend in the control ($\lambda_0 = \lambda_2 > 0$) with respect to the strength of the time trend in treatment arm 1 ($\lambda_1$) and according to the model used.

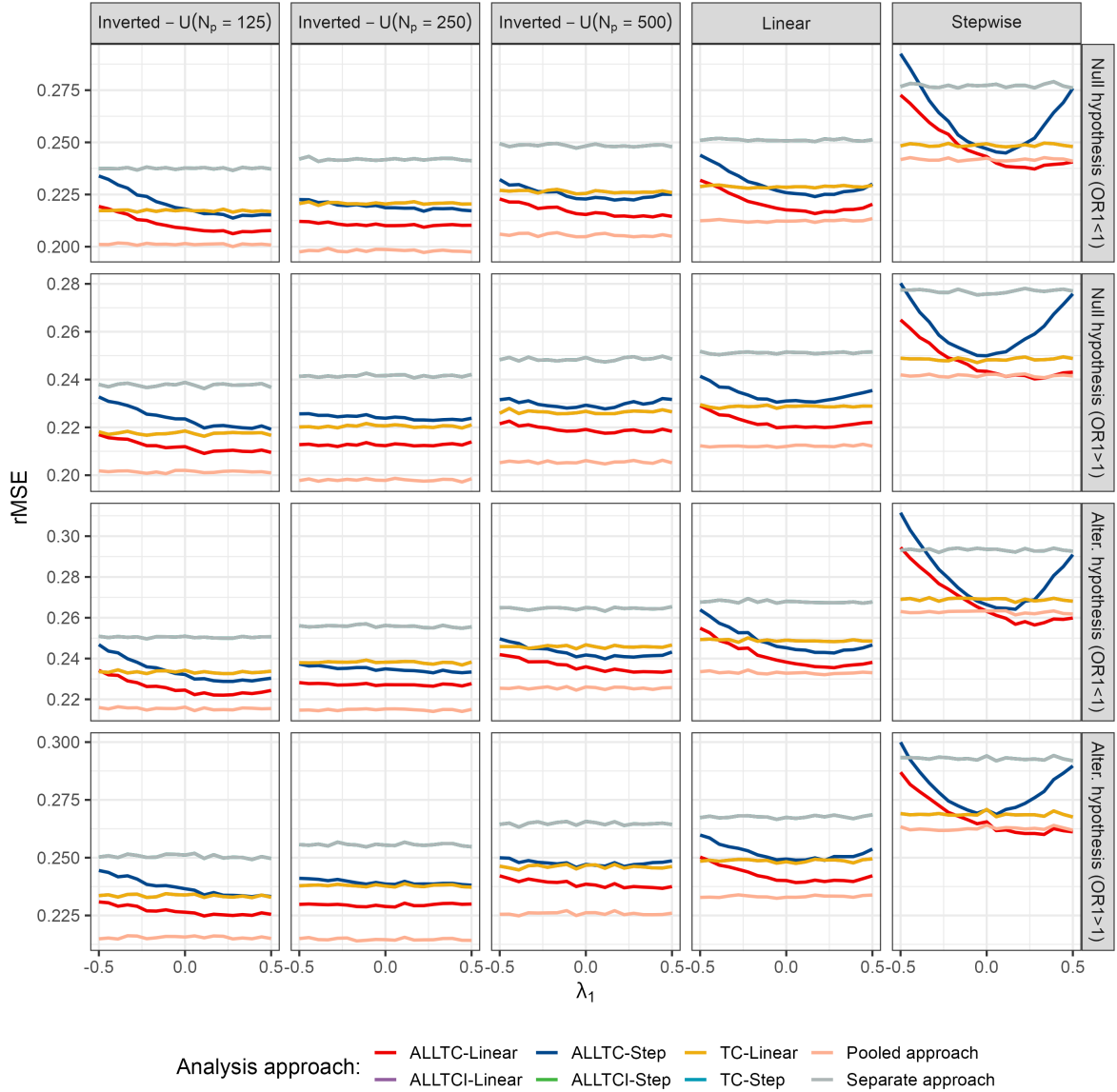

Figure S15: Root mean squared error of the treatment effect estimators ($\log(OR_2)$) for binary endpoints in the presence of linear, step-wise and inverted-U time trends when there is a positive time trend in the control ($\lambda_0 = \lambda_2 > 0$) with respect to the strength of the time trend in treatment arm 1 ($\lambda_1$) and according to the model used.

# F    Additional simulations

## F.1    Comparison of randomization procedures and patient entry time schemes

In this section, we compare different randomization procedures through simulations and discuss their role when incorporating non-concurrent controls. In particular, we consider two randomization procedures, simple randomization and block randomization [1], and study their impact on the type I error rate. In addition, we evaluate the impact of patient entry times when these are random rather than deterministic as considered in the article.

### F.1.1 Data generation

We simulated data of a two-period platform trial as described in Section 2 of the paper. In this case, however, we assumed that time trends are equal across groups and additive on the model scale such that the data are generated according to the model:

$$g\left(E(Y_j)\right) = \eta_0 + \sum_{k=1,2} \theta_k \cdot I(k_j = k) + f(t_j), \tag{6}$$

where $Y_j$, $g()$, $\eta_0$ and $\theta_{k_j}$ refer to the continuous or binary response, the link function (identity and logit functions for continuous or binary responses, respectively), the control response and treatment effects, respectively.

In this section, we considered a scenario in which there is no trend in the first period and the trend starts in the second period and is linear. Specifically, the time trend $f(t_j)$ is assumed to have the following pattern:

$$f(t_j) = \begin{cases} \lambda \frac{t_j - 1}{N - 1} & j > N_1 \\ 0 & j \le N_1, \end{cases} \tag{7}$$

where $N_1$ denotes the sample size in the first period and $N$ is the total sample size in the trial, so that the examined pattern corresponds to no time trend in the first period and linear time trend of strength $\lambda$ for all arms in the second period. Cases with moderate ($\lambda = 0.15$) and extreme ($\lambda = 5$) strengths of the time trend are examined.

For the patient entry times $t_j$, two options are discussed:

- deterministic entry times: $t_j = j$

- random entry times: $t_j \sim U(0, N_1/N)$ in the first period and $t_j \sim U(N_1/N, 1)$ in the second period

For deterministic entry times, the order of patients' entry in the trial is the same as that of the randomization sequence. In every unit of time, a patient enters into the platform. Thus, patients' index is equivalent to time at which patients enter in the study. Random entry times are uniformly distributed in both periods. For comparative purposes, random times are multiplied by the total number of sample size $N$ to achieve the same scale as in the case with deterministic entry times.

Furthermore, we assumed that the null hypothesis holds for treatment arm 2 (i.e. $\theta_2 = 0$ and $OR_2 = 1$). For continuous endpoints, we used control response $\eta_0 = 0$ and effect size for treatment 1 $\theta_1 = 0.25$. For binary endpoints, the control response rate was set to 0.3 and odds ratio for treatment 1 to 1.8. We considered a significance level $\alpha = 0.025$ to test the null hypothesis for treatment 2 against the one-sided alternative.

The ALLTC-Step model, i.e. model using all treatment data and control and adjusting for time by a step function, is used for the evaluation of the data. Each scenario was replicated 100.000 times.

### F.1.2 Randomization procedures

**Simple randomization per period.** The randomization procedure corresponds to the random allocation rule in which the sample size per arm in each period is prespecified. Thus, in each period, we take a random sample of the possible treatment allocation combinations given the sample sizes per arm and per period. So that this sequence represents the treatment allocation for each patient.

**Block randomization.** We considered the randomization procedure to be block randomization. Patients are assigned to arms following blocked randomization per period. In the first period (in which we have allocation ratio 1:1), we consider a block size of 4, while in the second period (in which we have 1:1:2) a block size of 12.

### F.1.3 Results

**Simple randomization per period.** For continuous endpoints, simple randomization maintains the type I error at 2.5% in the presence of moderate time trends for both, deterministic and random entry times. However, if the time trends have extreme strength, the type I error rate is inflated, reaching an empirical level of 3.3% for deterministic and random entry times.

In the setting with binary endpoints, the type I error rate is maintained, despite the strength of the time trend. That is, even when the time trend is extreme and regardless of whether the times were random or deterministic.

**Block randomization**. Similarly as in previous scenarios, for continuous endpoints when using block randomization, the type I error rate is maintained at 2.5% for fixed and random entry times whenever the time trend is moderate. In the presence of extreme time trends, however, this randomization procedure leads to strictly conservative results and empirical type I error rate of 0.7% for both, fixed and random entry times. For binary endpoints, the type I error rate is again maintained at 2.5% for moderate and extreme time trend strengths and deterministic and random entry times.

## F.2   Platform trials with three periods

In this section, we investigate the performance of the proposed models (1) and (2) in a trial with two treatment arms and a shared control in which arm 2 enters when the trial is ongoing. However, unlike the trial design reported in the article, here arm 1 finishes before arm 2 does. Hence, in this section, the platform trial is divided into three periods (instead of two periods as in the paper), where treatment arm 1 is active in the first 2 periods and treatment arm 2 in periods 2 and 3 (see Figure S16 for an illustration).

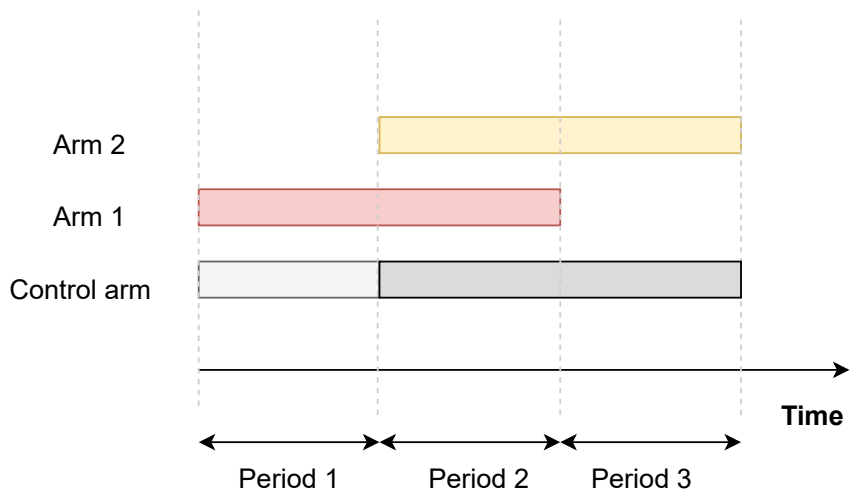

Figure S16: Scheme of a platform trial with 3 periods. Non-concurrent controls for arm 2 are shown in light grey.

### F.2.1   Data generation

We simulated data using the generating model (4) in the paper and following the same procedure than the one described in section F.1.1. We assumed linear time trend pattern, with the trend function $f(\cdot)$ given by:

$$f(j) = \lambda_{k_j} \frac{(j-1)}{(N-1)}$$

We investigated cases with equal strength of the time trend across all arms (i.e., $\lambda = \lambda_0 = \lambda_1 = \lambda_2$) as well as cases where the time trend in arm 1 differs (i.e., $\lambda_0 = \lambda_2 \neq \lambda_1$).

As in the main paper, we assumed that treatment 2 joins the platform after 1/2 of the patients have been allocated to treatment arm 1 and considered equal sample sizes $n$ in both treatment arms. In the newly considered scenarios, however, we used allocation ratio 1:1:1 in each period, thus treatment 1 leaves the trial earlier. Note that this results in larger sample size in the control group, as it is present in all 3 periods. Patients were assigned to arms according to block randomisation with block sizes of $2 \cdot (\#\text{active arms} + 1)$ in each period.

For continuous endpoints we assumed control response $\eta_0 = 0$, treatment effect for arm 1 of $\theta_1 = 0.25$ and treatment effects for arm 2 of $\theta_2 = 0$ or $\theta_2 = 0.25$ under the null and alternative hypotheses, respectively. In scenarios with binary endpoints, control response rate of $\eta_0 = 0.7$ and odds ratio for treatment 1 of 1.8 were used in all cases. Odds ratio for treatment 2 was set to 1 under the null hypothesis and to 1.8 under the alternative. We considered the sample size per treatment arm of $n = 210$, as this results in 80% power for the pooled analysis to detect the difference between the control group and treatment 2, assuming the given treatment effects, at 2.5% one-sided significance level in the case of no time trend.

### F.2.2 Results

The results of these additional simulations coincide with those presented in the main paper. In particular, under equal time trends, the power achieved using model (1) is higher than the power using model (2) and the separate approach, while the type I error is still controlled. If the time trend in treatment arm 1 differs from that in treatment arm 2 and control, the type I error rate is no longer maintained at 0.025. The type I error rate inflation, however, is less pronounced in the newly considered scenarios with three periods. This results from the fact that as there is now less overlap between the two treatments, treatment arm 1 contributes less to the estimation of the time trend and this estimation is then less affected by the different time trend in this arm.

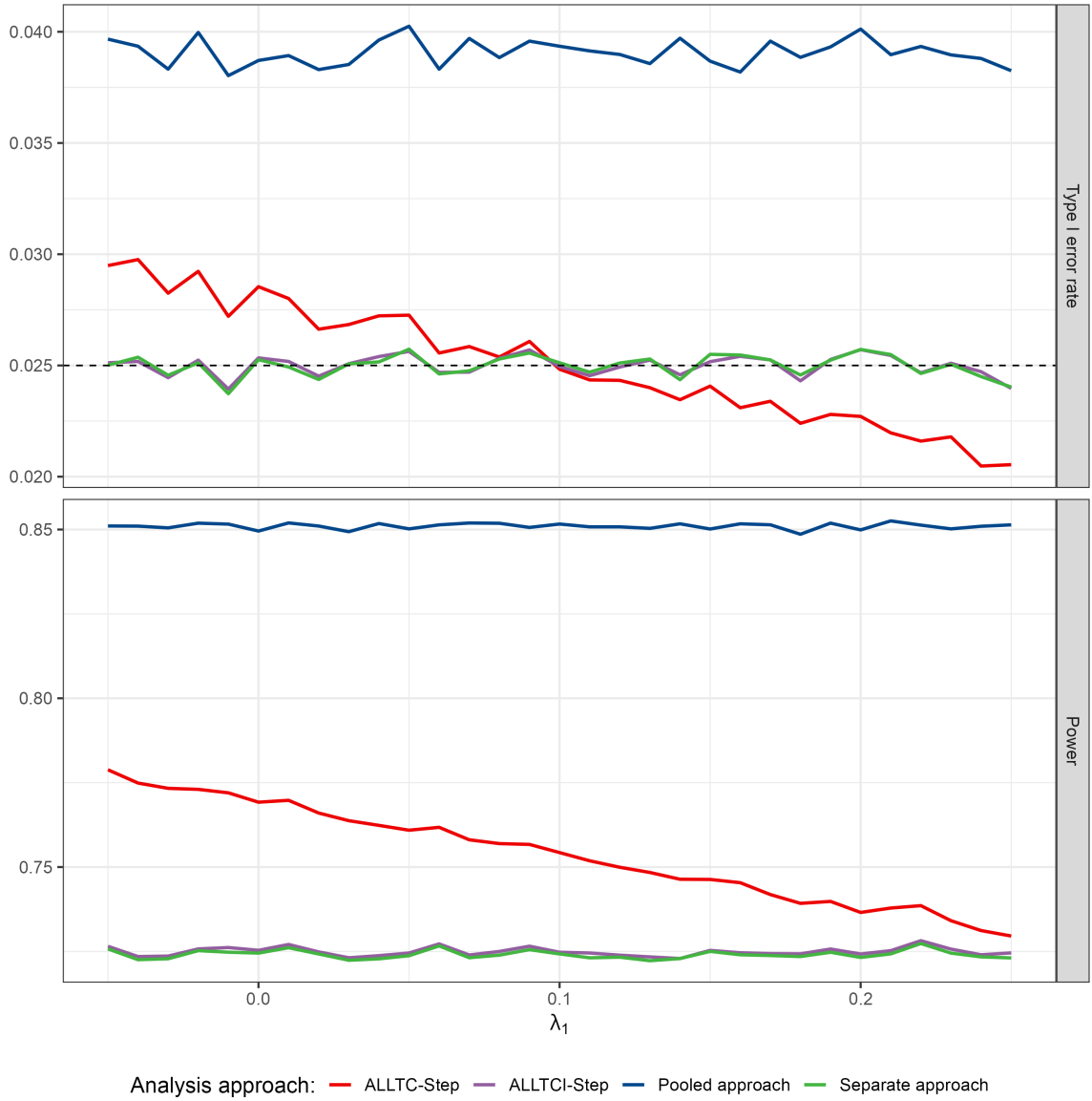

Figure S17: Type I error rate and power of rejecting $H_{02}$ for continuous endpoints in the presence of different linear time trends (for $\lambda_0 = \lambda_2 = 0.1$) with respect to the strength of the time trend in treatment arm 1 ($\lambda_1$) and according to the model used.

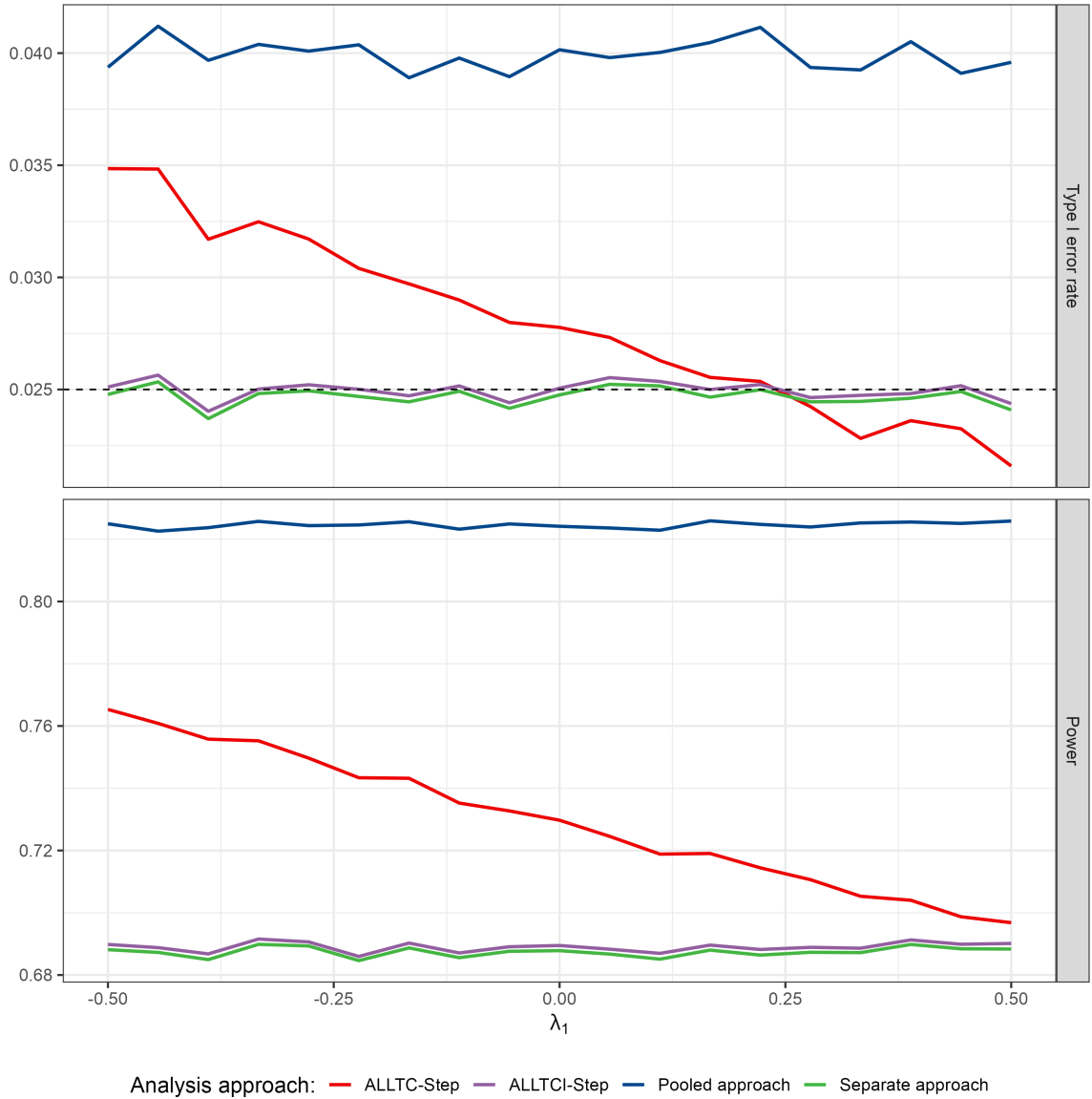

Figure S18: Type I error rate and power of rejecting $H_{02}$ for binary endpoints in the presence of different linear time trends when there is a positive time trend in the control ($\lambda_0 = \lambda_2 > 0$) with respect to the strength of the time trend in treatment arm 1 ($\lambda_1$) and according to the model used.



\clearpage

\end{document}